\documentclass[aps,prd,showpacs,showkeys,preprint,groupedaddress]{revtex4-1}

\usepackage{graphicx}
\usepackage{amsmath}
\usepackage{amssymb}
\usepackage{wrapfig}
\usepackage{indentfirst}

\begin{document}

\title{ Extracting the Odderon from $pp$ and ${p \bar p}$ scattering data }

\author{Andr\'as Ster}
\email[]{ster@rmki.kfki.hu}
\affiliation{Wigner Research Centre for Physics, Hungarian Academy of Sciences, H-1525 Budapest 114, P.O. Box 49, Hungary}

\author{L\'aszl\'o Jenkovszky}  
\email[]{jenk@bitp.kiev.ua}
\affiliation{BITP, National Academy of Sciences of Ukraine, Kiev, 03680 Ukraine} 
\affiliation{Wigner Research Centre for Physics, Hungarian Academy of Sciences, H-1525 Budapest 114, P.O. Box 49, Hungary} 

\author{Tam\'as~Cs\"org\H o}
\email[]{csorgo.tamas@wigner.mta.hu}
\affiliation{Wigner Research Centre for Physics, Hungarian Academy of Sciences, H-1525 Budapest 114, P.O. Box 49, Hungary}
\affiliation{KRF, H-3200 Gy\"ongy\"os, M\'atrai \'ut 36, Hungary}

\date{\today}

\begin{abstract}
Starting from a simple empirical parametrization of the scattering amplitude, successfully describing the dip-bump structure of elastic $pp$ scattering in $t$ at fixed values of $s$, we construct a toy model interpolating between missing energy intervals to extract the Odderon contribution from the difference between  ${\bar pp}$ and $pp$ elastic and total cross sections. The model is fitted to data from $\sqrt{s}$ = 23.5 GeV to 7 TeV and used 
to extract the Odderon and its ratio to the Pomeron. From our fits, a unit intercept Odderon follows, as predicted by J. Bartels, L.N. Lipatov, and G.P.~Vacca, on the basis of perturbative quantum chromodynamics.
\end{abstract}


\maketitle

\newpage
\section{Introduction} \label{s1}

The nature of the Odderon --- an asymptotic odd-$C$ Regge pole exchange, counterpart of the Pomeron --- for a long time remains a subject of debate. Although there is little doubt about its existence, we still lack direct evidence of the Odderon. Various reactions supposedly dominated by Odderon exchanges, called ``Odderon filters" \cite{filter}, may offer only indirect evidence either because of low statistics or contamination by competing exchanges. 

In quantum chromodynamics the Odderon corresponds to the exchange of an odd number of gluons. Relevant calculations were done in a number of papers; see \cite{Ewerz:2013kda} and earlier references therein.   

\begin{figure}\label{1}%
\includegraphics[width=10cm]{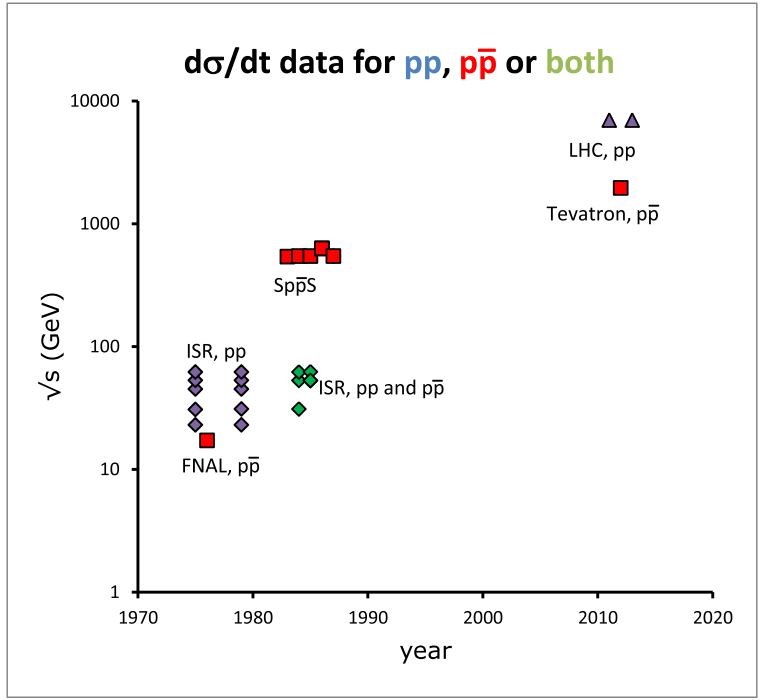}%
\caption{
Timeline of proton and antiproton elastic scattering measurements. 
New accelerators are run first at the maximum available energies;
however, at the start of the S$p\bar{p}$S accelerator, the $pp$ and the $p\bar{p}$ 
elastic scattering data were measured at the same $\sqrt{s} = $ 31, 53 and 62 GeV. 
}
\label{fig:accelarators}%
\end{figure}

The only direct way to see the Odderon is by comparing particle and antiparticle scattering at high enough energies. The high-energy proton-proton and proton-antiproton elastic scattering amplitude is a difference or sum of even and odd $C$-parity contributions, $A_{pp}^{\bar pp}(s,t)=
``Even" \pm ``Odd",$ where, essentially, the even part consists of the Pomeron and $f$ Reggeon, while the odd part contains the Odderon and the $\omega$ Reggeon. It is clear from the above formula that the odd component of the amplitude can be 
extracted from the difference of the proton-antiproton and proton-proton scattering amplitudes, and, since at high enough energies the contributions from secondary Regge trajectories die out, this difference offers a direct way of extracting the Odderon contribution. Unfortunatelly, $pp$ and $\bar pp$ elasctic scatterings were typically
measured at different $\sqrt s$, with the exception of the ISR energies of 31, 53, 62 GeV (see Fig. \ref{fig:accelarators}).

At present, the only way to extract the Odderon from the difference of the ${\bar pp}$ and $pp$ scattering amplitudes is by means of a reliable interpolation of both amplitudes (or cross sections) over the missing energy regions. While the energy dependence of the forward amplitude (or total cross sections) is controlled by the Regge pole theory, its $t$ dependence, especially in the dip-bump region, to a large extent is model dependent and  unpredictable. 

A simple, general, and reliable parametrization of the complicated diffraction structure at high energies at any fixed energy is a sum of two exponentials in $t$ related by a  complex phase $e^{i\phi}.$   
Using this generic expression, Phillips and Barger (PB)~\cite{Phillips:1974vt} (for brevity we shall refer to it as the PB ansatz) obtained good fits to the proton-proton differential cross sections, including the dip-bump region at several CERN ISR fixed energies. In Refs. \cite{Grau:2012wy, Fagundes:2013aja, Fagundes:2013cja} the model was extended and improved, in particular, by accounting for the nonexponential behavior in the low $|t|$ region. In Ref. \cite{JL} the parametrization was shown to be applicable to proton-antiproton scattering as well. Its connection with inelastic reactions is discussed on p. 185 of Ref. \cite{Goulianos:1982vk}.  

The PB ansatz reads
\begin{equation}\label{Barger}
{\cal A}(s,t)=i[\sqrt A\exp(Bt/2)+\exp(i\phi(s))\sqrt C\exp(Dt/2)],
\end{equation}
where $s$ and $t$ are the standard Mandelstam variables; $A, B, C, D,$ and $\phi$ were fitted to each energy independently;
i.e., energy dependence in the PB ansatz enters parametrically. 

The success of the simple PB parametrization motivates its further improvement, extension, and 
utilization. In papers \cite{Grau:2012wy, Fagundes:2013aja} the low-$|t|$ behavior of the 
PB ansatz was improved by modifying its simple exponential behavior by (a) inclusion of 
a two-pion threshold required by analyticity and 
(b) by means of a multiplicative factor reflecting the proton form factor. 
Achieving good fits to the TOTEM data \cite{Antchev:2011zz}, 
 at the LHC energies of $\sqrt{s}=7$ TeV, the modified ansatz was used to predict the behavior of the observables at future energies as well as the expected asymptotic behavior of the cross sections.

The model was also tested in \cite{Fagundes:2013cja, JL} against the TOTEM data \cite{Antchev:2011zz}, 
and, contrary to many alternative models, it works reasonably well. 
In Ref. \cite{Fagundes:2013aja} the PB ansatz Eq. (\ref{Barger}) was improved; in particular, its small-$t$ behavior was modified.  The fitted parameters are collected in Table II of that paper.  
        For example, the authors of \cite{Fagundes:2013aja}
quote the following values for these parameters for $pp$ scattering at 
$\sqrt{s}=53$ GeV: $\sqrt{A}=6.55,\ \ \sqrt{C}=0.034$ in $\sqrt{{\rm mb/GeV}^2}$ and $B=10.20,\  D=1.7$ in GeV$^{-2},\ \phi=2.53\ {\rm rad}$.
 
         As already mentioned, in addition to elastic $pp$ scattering, the PB ansatz also describes $p\bar{p}$
 data, with a different set of the parameters (see below), thus opening the way to be used as a 
tool in extracting the Odderon from the difference of the two. 
However, in its original form, the PB ansatz does not describe the $\sqrt{s}$ dependence of the model 
parameters.

In this paper, we try to remedy this limitation by combining the appealingly simple and efficient form of its $t$ dependence with energy dependence inspired by the
Regge pole model. Work in this direction was started in papers \cite{Grau:2012wy, Fagundes:2013aja, Fagundes:2013cja, JL, Jenkovszky:2014bwa}.

We address the following issues: (1) we smoothly interpolate between the values of the parameters
fitted at fixed energy values, (2) extract the Odderon contribution from the 
difference of the $\bar pp$ and $pp$ cross sections, and (3) compare the energy dependence of this difference with the  prediction \cite{BLV} based on perturbative quantum chromodynamics, by which the intercept of the Odderon trajectory equal to one.  
  
The structure of this paper is as follows: in the next section, we introduce the $\sqrt{s}$ dependence of the parameters of the PB model, which in the next section are determined from the fits to data. Then we discuss the results, in particular the $\sqrt{s}$ dependence of the results, including, for example, the Odderon contribution and the Odderon/Pomeron ratio. Finally, we summarize and conclude.

\section{The generalized PB model}

We use the norm where
\begin{equation}\label{norm1}
\sigma_{tot}=4\pi \Im A(t=0)=4\pi[\sqrt{A}+\sqrt{C}\cos{\phi}]
\end{equation}
and 
\begin{equation}\label{Lia2}
\frac{d\sigma}{dt}=\pi |{\cal A}(t)|^2= \pi[Ae^{Bt}+Ce^{Dt}+2\sqrt{A}\sqrt{C}e^{(B+D)t/2}\cos{\phi}].
\end{equation}

Following the Regge pole theory, we make the following assignment,
\begin{equation} \label{AC}
\sqrt{A}\rightarrow \sqrt{A(s)}=a_1s^{-\epsilon_{a_1}}+a_2s^{\epsilon_{a_2}},\ \
\sqrt C \rightarrow \sqrt{C(s)}=cs^{\epsilon_{c}}
\end{equation} 
inspired by the Donnachie and Landshoff model  \cite{Donnachie:1983hf} of cross sections [see Eq. (\ref{norm1})] with effective falling (subleading Reggeons) and rising (Pomeron) components. It follows from our fits that the falling (subleading Reggeon) components in $\sqrt C$ are small; hence, they are neglected.

The slopes $B$ and $D$ in the Regge pole theory are unambiguously logarithmic in $s$, providing shrinkage of the cone:
\begin{equation}\label{BD}
B\rightarrow B(s)=b_0+b_1\ln(s/s_0),\ \ \ D\rightarrow D(s)=d_0+d_1\ln(s/s_0).
\end{equation}
In the above formulae a normalization factor $s_0=1$ GeV$^2$ is implied. 

The phase $\phi$ is the weakest point of this ``toy" model or generalized PB model. In Regge theory, it should depend on $t$ rather than on $s$. Fortunately, at high $\sqrt s$ the dependence of $\phi$ on
energy is weak (see the fits below). However, this is not the case as the energy decreases. The best we can do is to fit the data with
\begin{equation}\label{phi}
cos(\phi(s)) = k_0+k_1s^{-\epsilon_{cos}}.
\end{equation}  
The ``low"-energy behavior is a weak point in any case. Apart from the varying phase, we must account in some way for the subleading ($f$ and $\omega$) Reggeon contributions. This is done partly by the inclusion in $\sqrt{A(s)}$ of
a decreasing term (absent in $\sqrt C$). A complete treatment of these terms with proper $t$-dependent signatures will require a radical revision of the model, and we hope to come back to this issue in the future. 

Now we proceed with this simple approach that has a chance to be viable at high energies, where the Pomeron and Odderon dominate \cite{Jenkovszky:2011hu} and the above complications may be insignificant.    

  To understand better the existence of any connection between the ansatz (\ref{Barger}) and the Regge pole model,
we plot the values of the parameters $A, B, C, D,$ and $\phi$ against $s$ and fit their ``experimental" values
to Regge-like formulas. 

This can be done in two complementary ways: A successive ``two-step" fit. First, we acquire the values of the parameters $A,\ B,\ C,\ D,\ \phi$ from the fits to the $pp$
and $\bar pp$ data, then we fit their Regge forms (see below) to the obtained ``experimental" values of  $A,\ B,\ C,\ D,\ \phi$.
Alternatively, one may determine the parameters of Eqs. (\ref{AC}) and (\ref{BD}) 
from a single simultaneous fit to all available data.
We chose the first option (5 parameters) since otherwise there were too many (at least 12) 
free parameters. Thus, we proceed with a two-step fit, by which the final values are 
determined from a fit to the ``experimental" values of $A, B, C, D,$ and $\phi$.

We fitted separately $pp$ and $p \bar p$ in two variables, $s$ and $t,$ 
by using $pp$ and $p \bar p$ data on total and differential cross sections 
ranging from the ISR  to the LHC  for $pp$ and from S$p\bar p$S to the Tevatron for $\bar pp$. 

Here the following remarks are in order:

(1) It is clear from Eqs. (\ref{AC}) and (\ref{BD}) that, while the parameters $A$ and $C$ are particularly sensitive to the data on  total cross sections, $B$ and $D$ are correlated mainly with the differential cross sections (the slopes).

(2) Although we are interested mainly in the high-energy behavior (the Odderon), low-energy effects cannot be fully neglected. They are taken into account approximately by including in $A$ and $C$ subleading terms of the type $s^{\epsilon},\ \epsilon\approx 0.5$.

(3) Having fitted $A, B, C, D,$ and $\phi,$ we perform a cross-check by calculating the resulting total cross sections.

(4) At high energies, the proton-proton and antiproton-proton total cross sections are supposed to converge. 
(We consider only this simple option, although we are aware of alternatives.) 
Since the existing data are not yet in this asymptotic domain,
we introduce an extra constraint. 

(5) The most delicate issue is the phase, which in Regge phenomenology
is expected to be $t$, rather than $s$, dependent. 
Our fits (and those of \cite{Grau:2012wy, Fagundes:2013aja, Fagundes:2013cja})
show considerable energy dependence of the phase at low energies but weak dependence at high energies, 
where we are particularly interested in looking for the Odderon signal. 
Postponing the introduction of a true Regge-pole-motivated, $t$-dependent phase to a further study,
here we assume a simple parametrization $\cos\phi=k_0+k_1s^{\epsilon_{\phi}}$.

\section{Fitting the model to the data}

We have calculated the $s$ dependence of the parameters using a fitting strategy consisting of three consecutive steps described in detail below. In doing so, the following criteria were 
applied:

\begin{itemize}

\item  best $\chi^2$ for each fit;

\item  the $-t$ range was set within $~0.35$-$2.5$ GeV$^2$;

\item  for each fixed energy, the model was fitted simultaneously to $d\sigma/dt$ and $\sigma_{tot}$
(focusing on the dip region); 

\item  from the resulting fits, $\sigma_{tot}$ was reconstructed in the whole available energy range.


\end{itemize}

The values of the fitted parameters, in general, are consistent, within about 30\%, with those obtained in 
Refs.~\cite{Phillips:1974vt,Fagundes:2013aja}, except for the small value of $\sqrt{C}$ deviating considerably from
those quoted in the above papers, which used a different $t$ range for their data analysis. 

\clearpage

\subsection{Step 1: Fitting the parameters $A, B,...$ to $d\sigma/dt$ and $\sigma_{tot}$ for fixed energies}

\begin{figure}\label{2}%
\includegraphics[width=7cm]{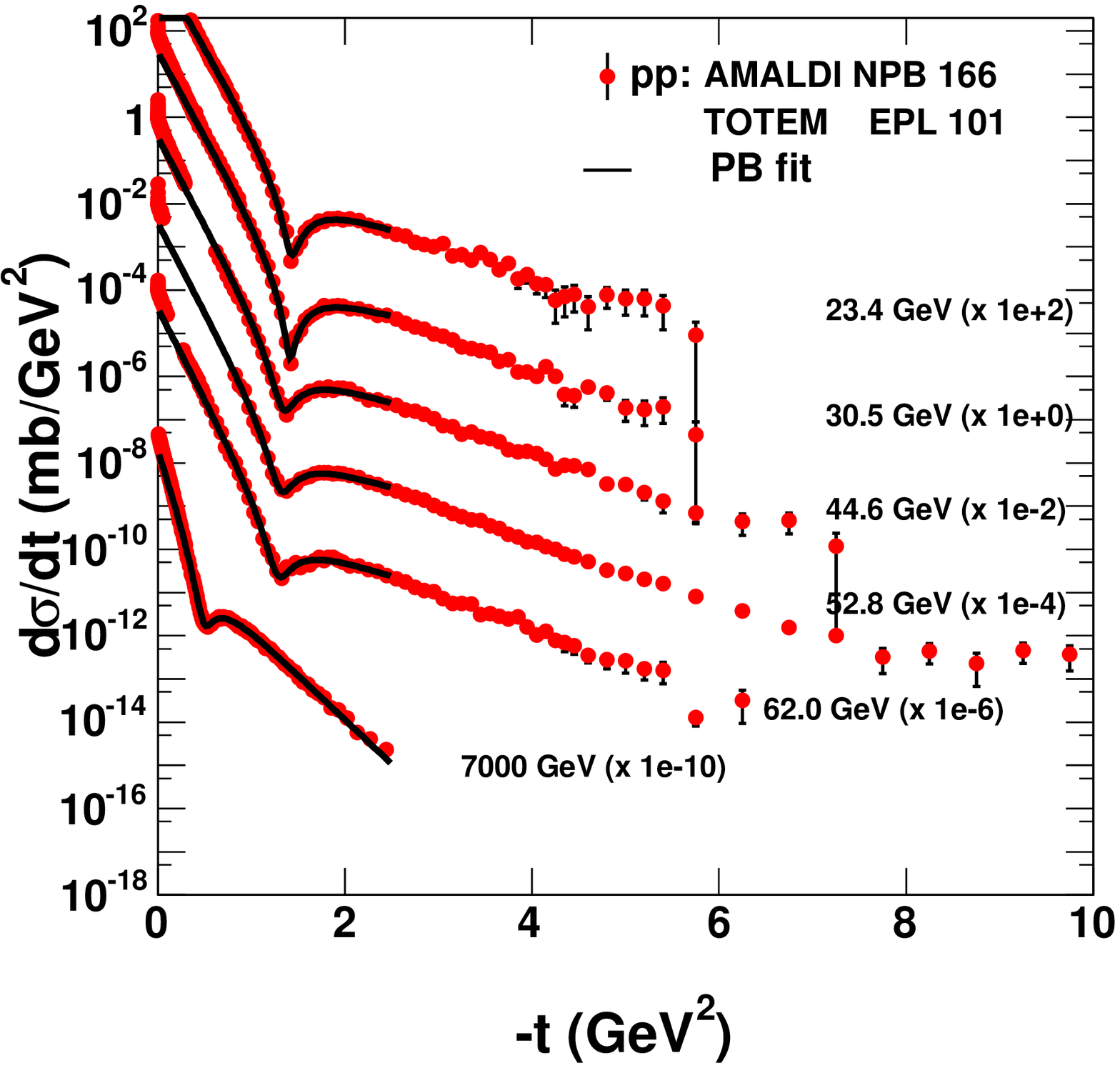}%
\includegraphics[width=7cm]{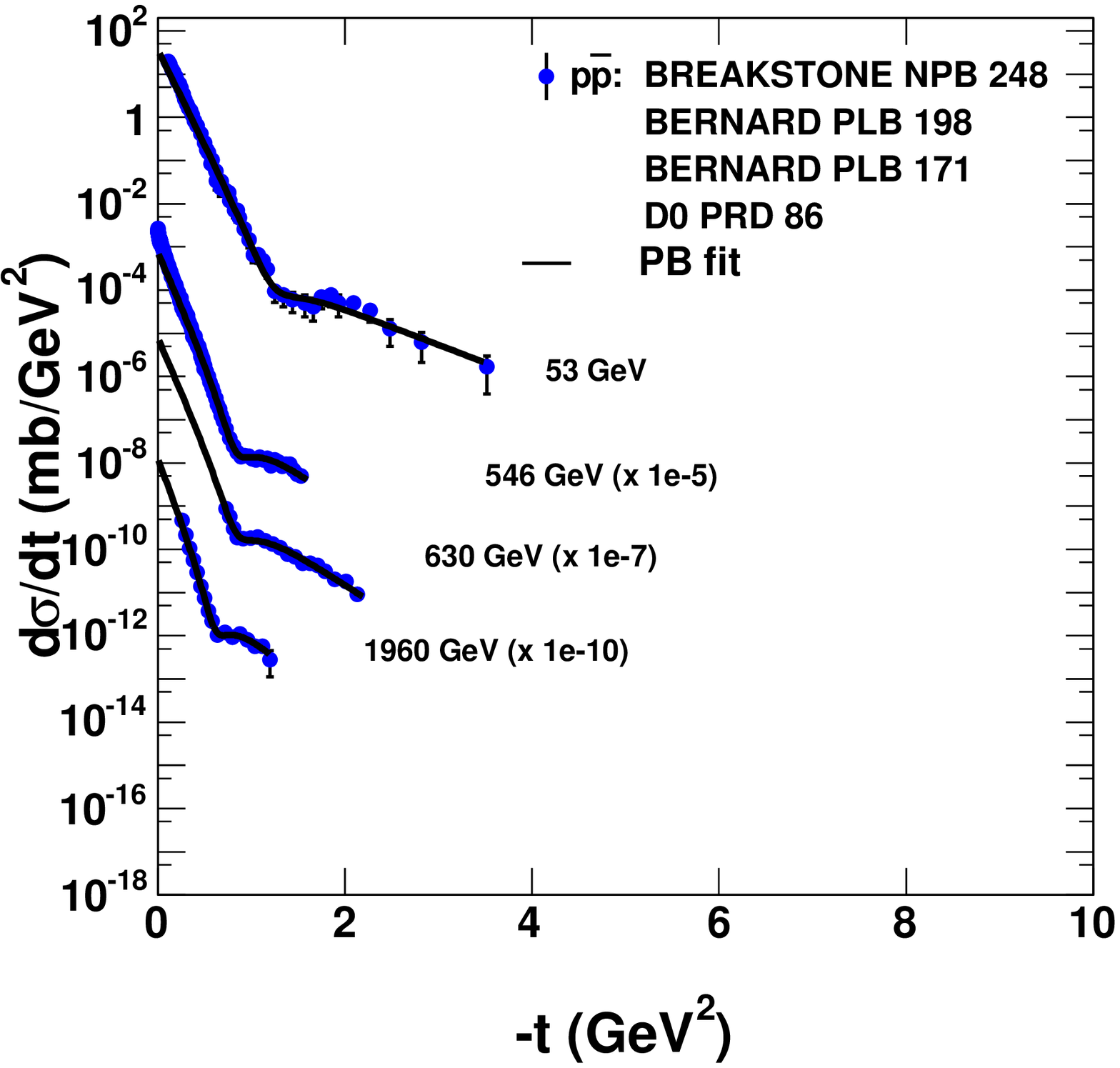}%
\linebreak
\includegraphics[width=7cm]{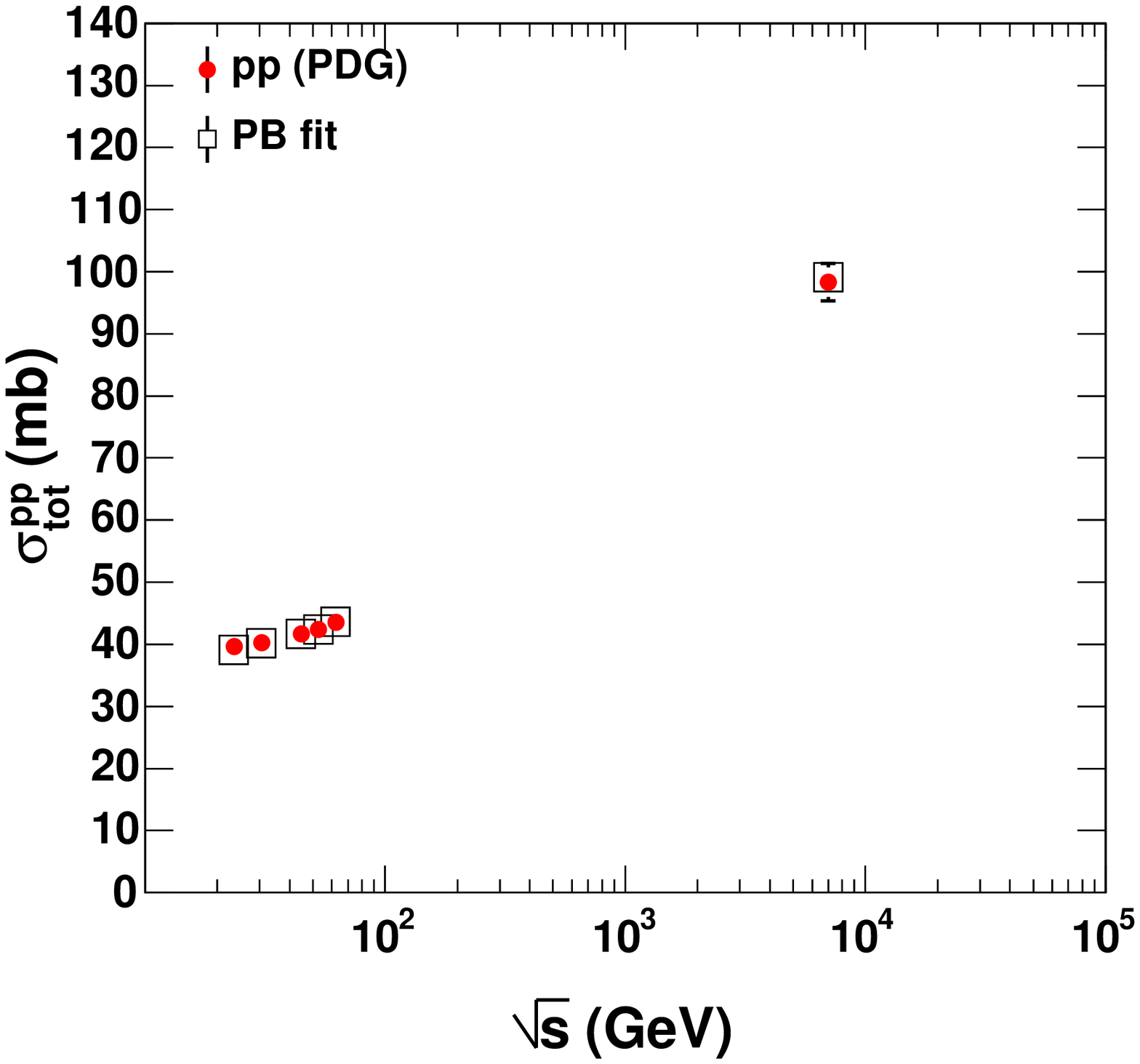}%
\includegraphics[width=7cm]{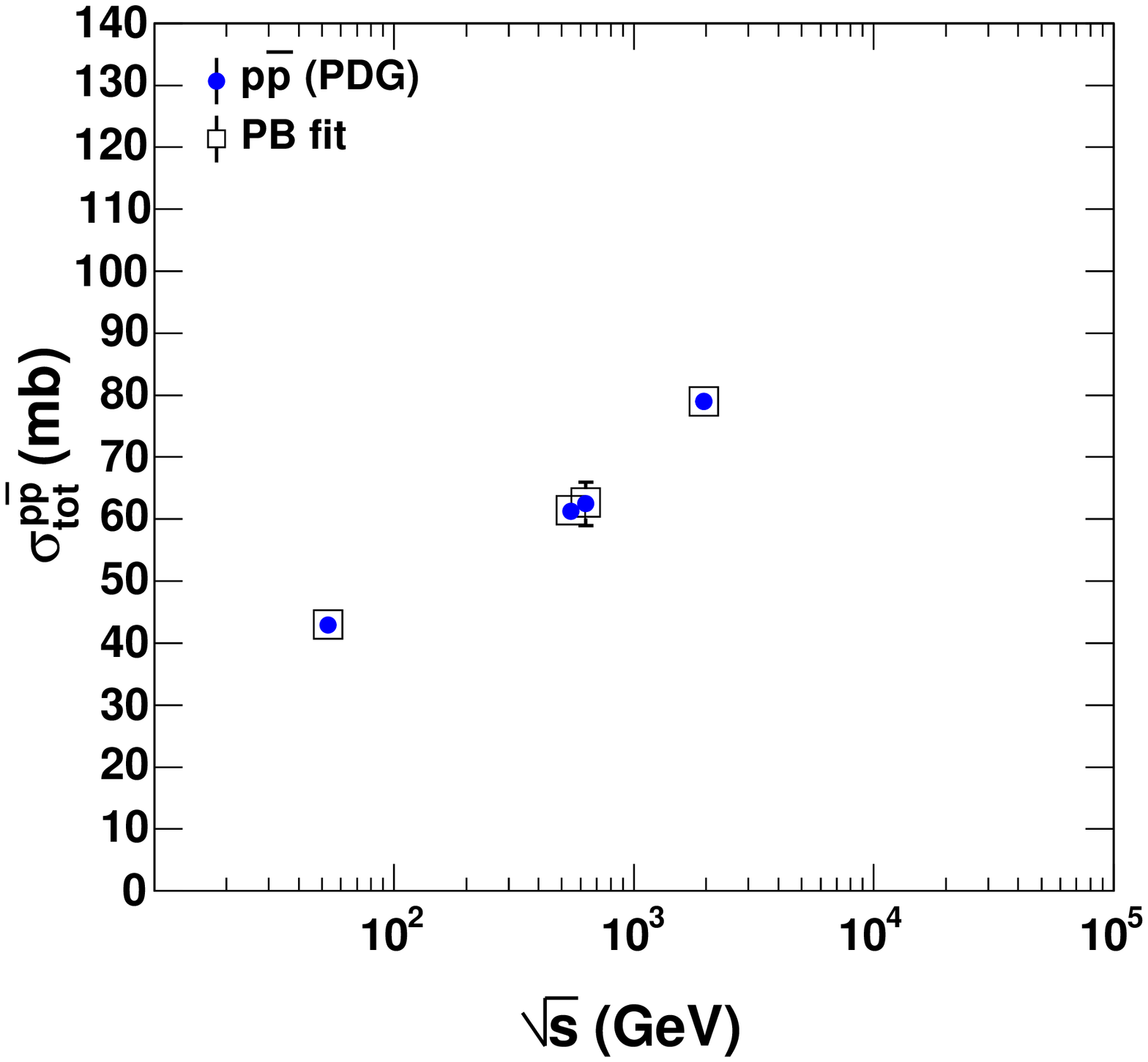}%
\caption{The PB model fitted to the pp and $\bar{p}p$ data at discrete energy values. 
The $d\sigma/dt$ data are from 
Refs.~\cite{Amaldi:1979kd,Antchev:2013gaa,break:npb248,Bernard:1987vq,Bernard:1986ye,Abazov:2012qb}.
The data on $\sigma_{tot}$ are from the Particle Data Group database~\cite{Beringer:1900zz}.}
\label{fig:BP_fits}%
\end{figure}

Figure \ref{fig:BP_fits} shows a fit to the data on the $pp$ and $\bar pp$ differential and total cross sections. The parameters $A, B, C, D,$ and $\phi$ were fitted to each energy separately. Given the simplicity of the model, the fits look reasonable. 

Figure \ref{fig:pp_params1} shows the fitted values of the parameters $A, B, C, D,$ and $\phi$ both for $pp$ and $\bar pp$ scattering to be used as ``experimental" data in the second stage of our fitting procedure, in which the explicit expressions (\ref{AC}), (\ref{BD}) and (\ref{phi}) are inserted. The fitted values of the parameters and relevant $\chi^2/NDF $ values are quoted in Tables \ref{tab:results1} and \ref{tab:results2}.

Note the difference between the present model and that of Refs. \cite{Grau:2012wy, Fagundes:2013aja, Fagundes:2013cja} in the low-$|t|$ behavior of the differential cross section, corrected in \cite{Grau:2012wy, Fagundes:2013aja, Fagundes:2013cja, Pancheri:2014roa} and taking account of the deviation from an exponential. Normalization in Ref. \cite{Phillips:1974vt} is arbitrary since, in that paper, only the differential cross section is shown. Total cross sections are not shown in the above papers.

\begin{table}[t]
\begin{center}
\begin{tabular}{|r|l|l|l|l|l|c|}
\hline
\hline
Energy &$\sqrt{A}$  &B  &$\sqrt{C}$ &D  &$cos(\phi)$ &$\chi^{2}$/NDF \\
(GeV)  &            &   &           &   &       & \\
\hline
  23.4& 3.13$\pm$ 0.6\%& 8.66$\pm$ 0.4\%& 0.019$\pm$ 8.3\%& 1.54$\pm$ 5.1\%& -0.97$\pm$ 0.3\%& 1.6 \\
  30.5& 3.21$\pm$ 0.2\%& 8.95$\pm$ 0.3\%& 0.014$\pm$ 7.4\%& 1.28$\pm$ 5.6\%& -0.98$\pm$ 0.2\%& 1.1 \\
  44.6& 3.33$\pm$ 0.7\%& 9.32$\pm$ 0.5\%& 0.017$\pm$ 8.0\%& 1.45$\pm$ 5.3\%& -0.93$\pm$ 0.8\%& 1.7 \\
  52.8& 3.38$\pm$ 0.3\%& 9.44$\pm$ 0.6\%& 0.017$\pm$ 7.6\%& 1.43$\pm$ 5.0\%& -0.92$\pm$ 0.9\%& 1.1 \\
  62.0& 3.49$\pm$ 0.5\%& 9.66$\pm$ 0.6\%& 0.018$\pm$ 9.9\%& 1.53$\pm$ 6.3\%& -0.92$\pm$ 1.6\%& 1.5 \\
7000.0& 8.51$\pm$ 1.6\%&15.05$\pm$ 0.8\%& 0.670$\pm$ 2.3\%& 4.71$\pm$ 0.8\%& -0.93$\pm$ 0.3\%& 1.4 \\
\hline
\hline
\end{tabular}
\end{center}
\caption {Values of the parameters from a fit to the $pp$ data at various $\sqrt{s}$. The quoted errors correspond to the relative errors, as given by CERN MINUIT fitting package (status= converged, error matrix accurate). }
\label{tab:results1}
\end{table}

\begin{table}
\begin{center}
\begin{tabular}{|r|l|l|l|l|l|c|}
\hline
\hline
Energy &$\sqrt{A}$  &B  &$\sqrt{C}$ &D  &$cos(\phi)$ &$\chi^{2}$/NDF \\
(GeV)  &            &   &           &   &       & \\
\hline
  63& 3.43$\pm$ 1.1\%&10.07$\pm$ 1.3\%& 0.022$\pm$30.8\%& 1.90$\pm$14.8\%& -0.60$\pm$22.7\%& 0.7 \\
 546& 5.06$\pm$ 1.2\%&11.25$\pm$ 1.3\%& 0.204$\pm$21.0\%& 3.55$\pm$ 8.6\%& -0.86$\pm$ 2.7\%& 0.6 \\
 630& 5.13$\pm$ 3.9\%&11.26$\pm$ 3.7\%& 0.176$\pm$26.6\%& 3.23$\pm$ 9.6\%& -0.81$\pm$ 7.9\%& 0.5 \\
1960& 6.85$\pm$ 3.7\%&12.46$\pm$ 3.3\%& 0.629$\pm$41.6\%& 4.69$\pm$15.4\%& -0.90$\pm$ 3.6\%& 0.4 \\
\hline
\hline
\end{tabular}
\end{center}
\caption {Values of the parameters fitted to $p \bar p$ data}
\label{tab:results2}
\end{table}

\pagebreak

\subsection{Step 2: Fitting the 
parameters $a_i,\ b_i,...$ entering Eqs. (\ref{AC}), (\ref{BD}) and 
(\ref{phi}) to the ``data" $A,\ B,\ C,\ D,\ cos(\phi)...$, quoted in Tables I and II.}
 
The resulting values of the parameters after the second stage of fitting are

\begin{eqnarray}
\sqrt{A_{pp}(s)}&=& 1.31 s^{0.106} + 3.90 s^{-0.298}, \\
\sqrt{C_{pp}(s)}&=& 0.00117 s^{0.358},\nonumber\\
B_{pp}(s)&=&5.13 + 0.555 \ln s,\nonumber \\
D_{pp}(s)&=&-0.838 + 0.312 \ln s. \nonumber \\
cos(\phi_{pp}(s))&=& -0.928 - 0.863 s^{-0.429}. \nonumber
\label{eq:PBfunc_pp}
\end{eqnarray}

\begin{eqnarray}
\sqrt{A_{p\overline{p}}(s)}&=& 1.31 s^{0.106} + 4.28 s^{-0.298}, \nonumber\\
\sqrt{C_{p\overline{p}}(s)}&=& 0.00177 s^{0.358},\nonumber\\
B_{p\overline{p}}(s)&=&7.87 + 0.274 \ln s,\nonumber \\
D_{p\overline{p}}(s)&=&-0.552 + 0.312 \ln s, \nonumber \\
cos(\phi_{p\overline{p}}(s))&=& -0.928 + 4.37 s^{-0.328}. \nonumber
\label{eq:PBfunc_ppa}
\end{eqnarray}

\clearpage
\begin{figure}[ht]\label{3}%
\includegraphics[width=7cm]{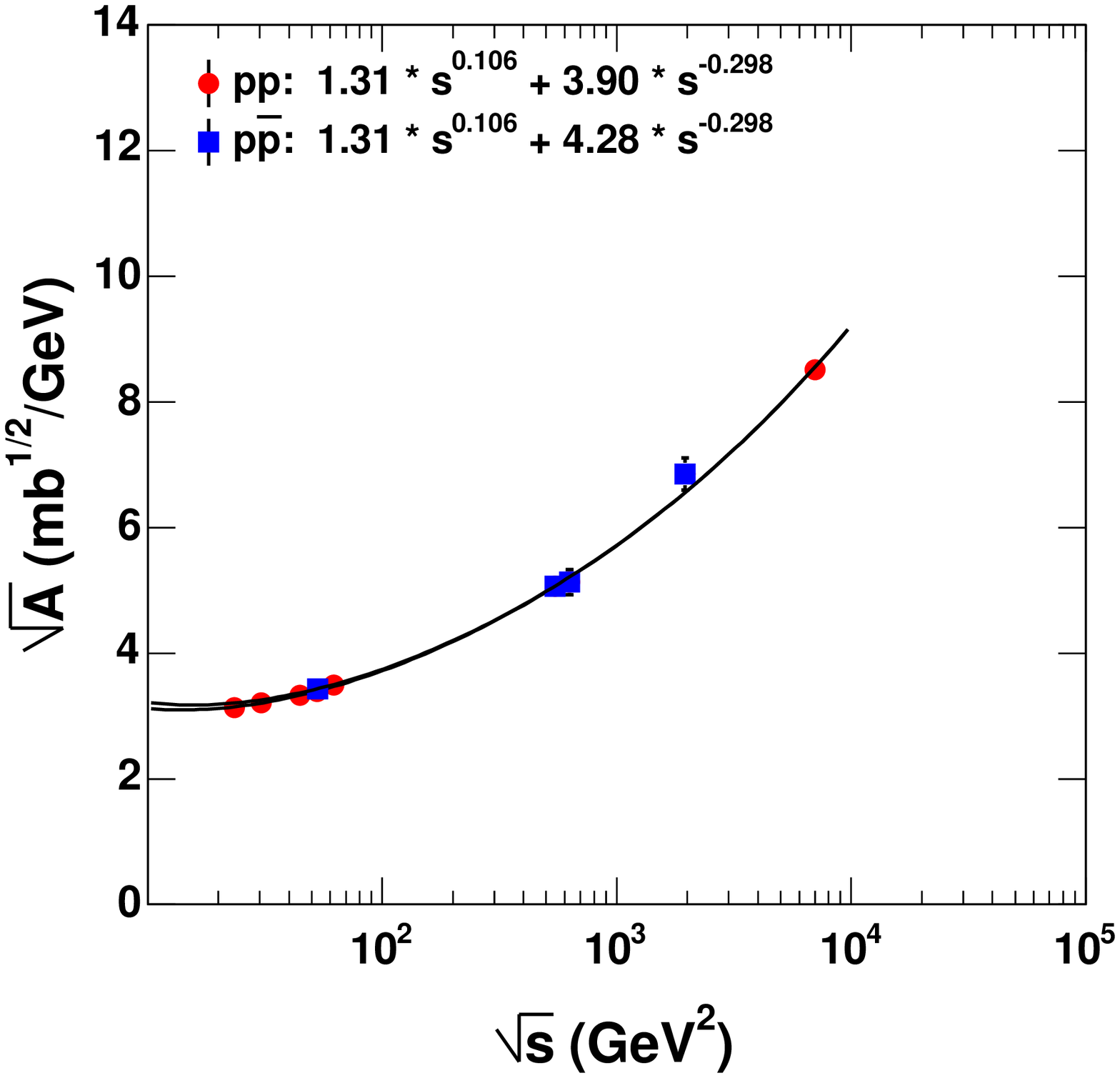}%
\includegraphics[width=7cm]{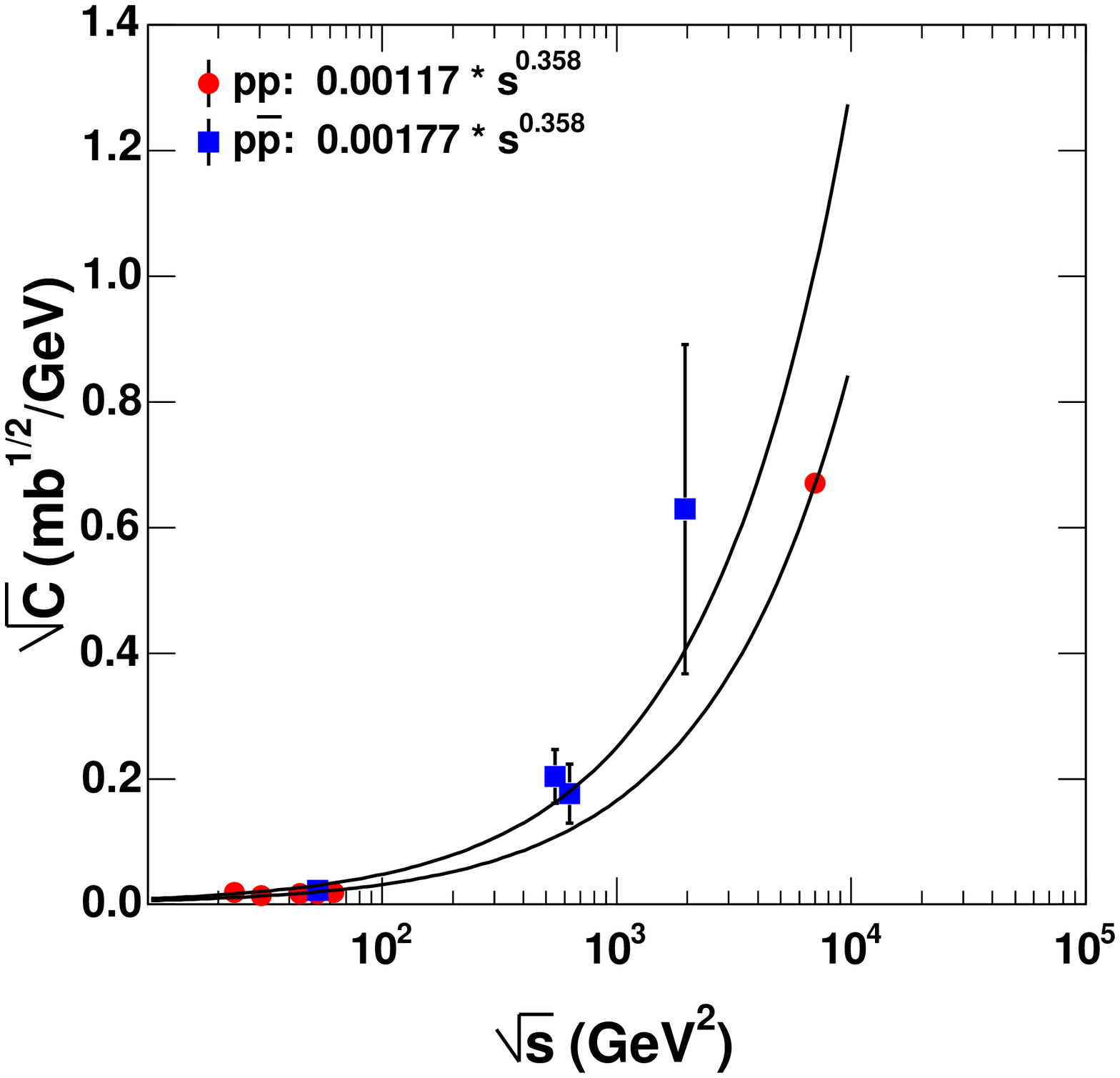}%
\linebreak
\includegraphics[width=7cm]{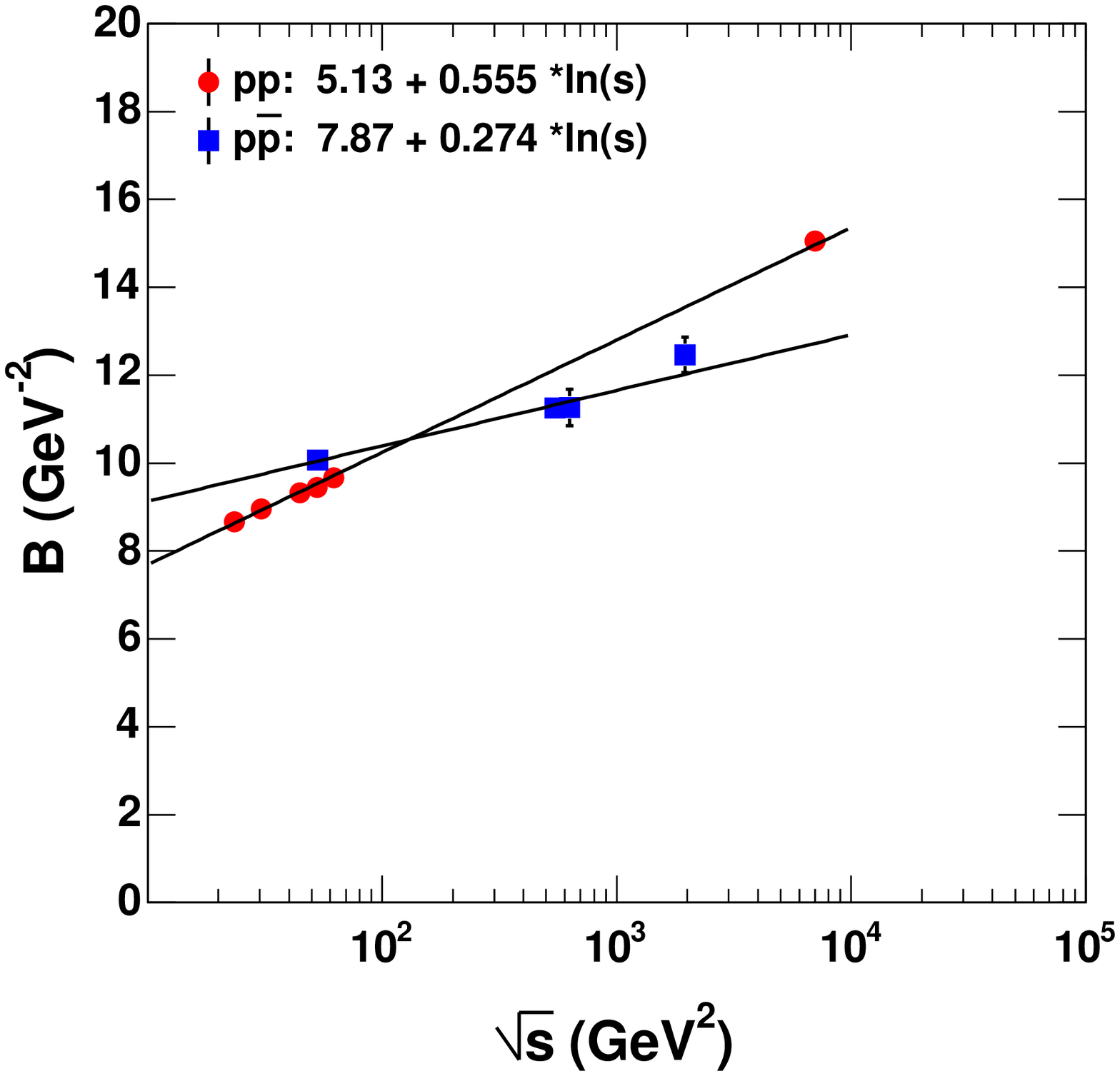}%
\includegraphics[width=7cm]{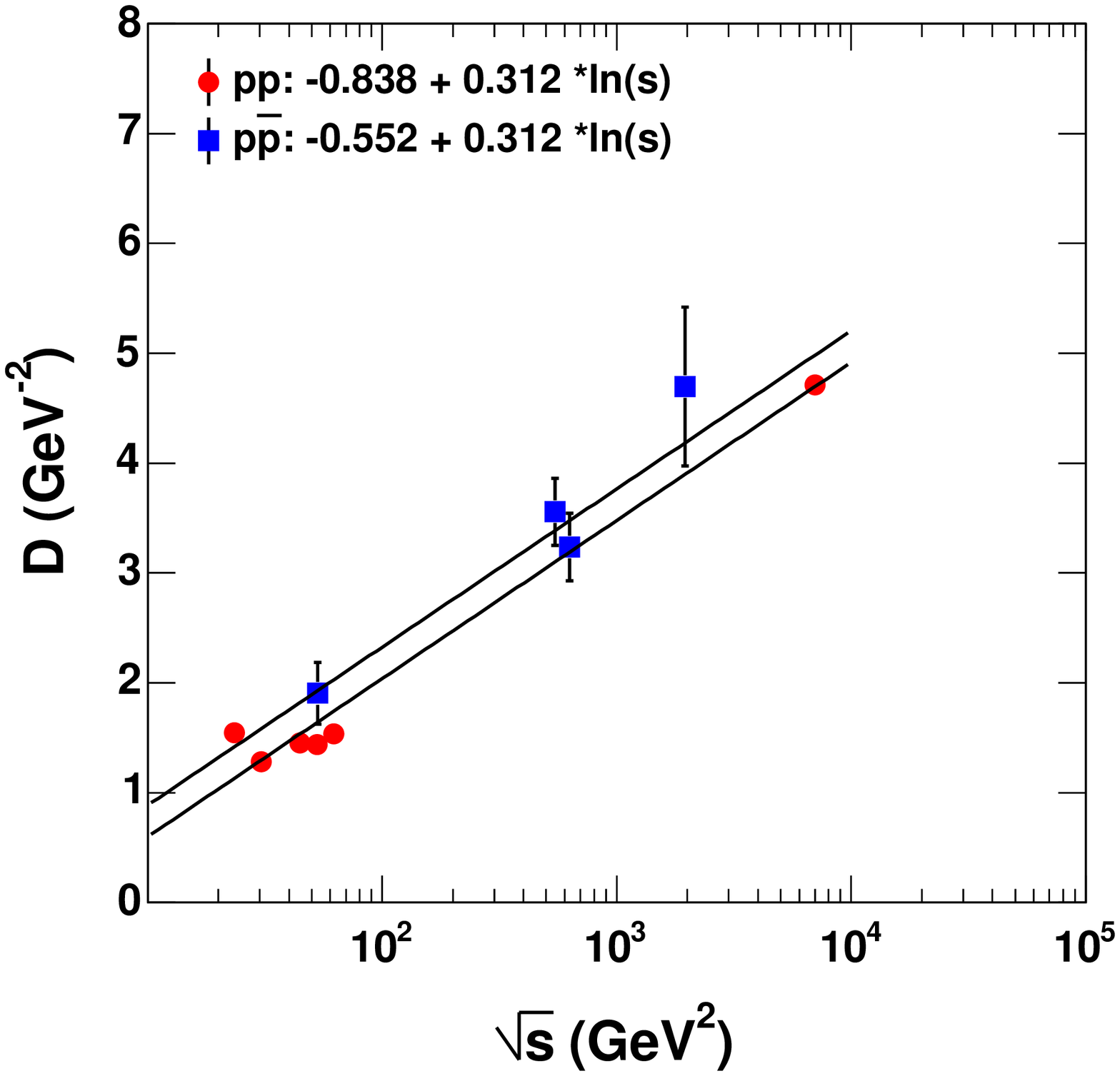}%
\linebreak
\includegraphics[width=7cm]{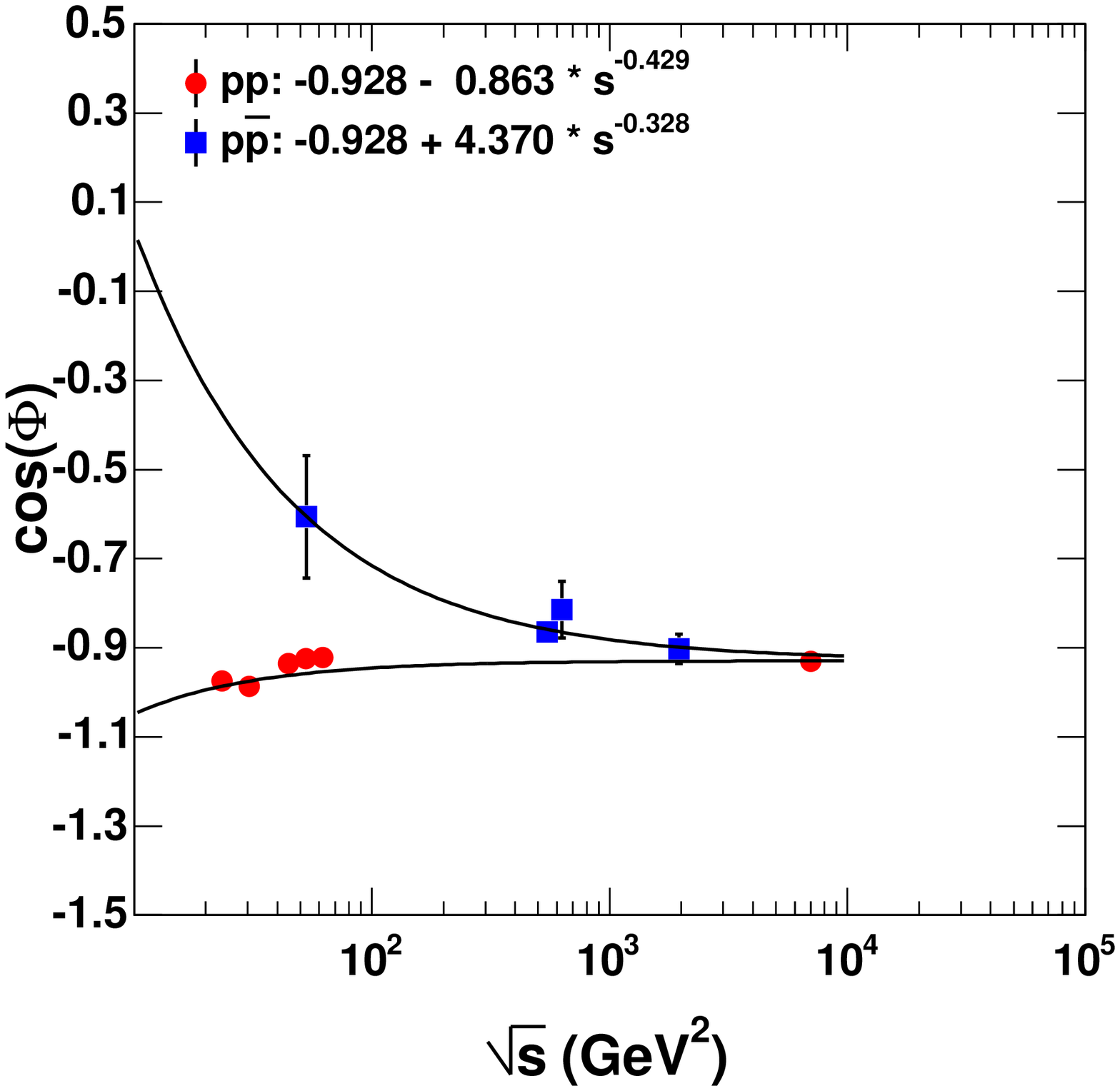}%
\includegraphics[width=7cm]{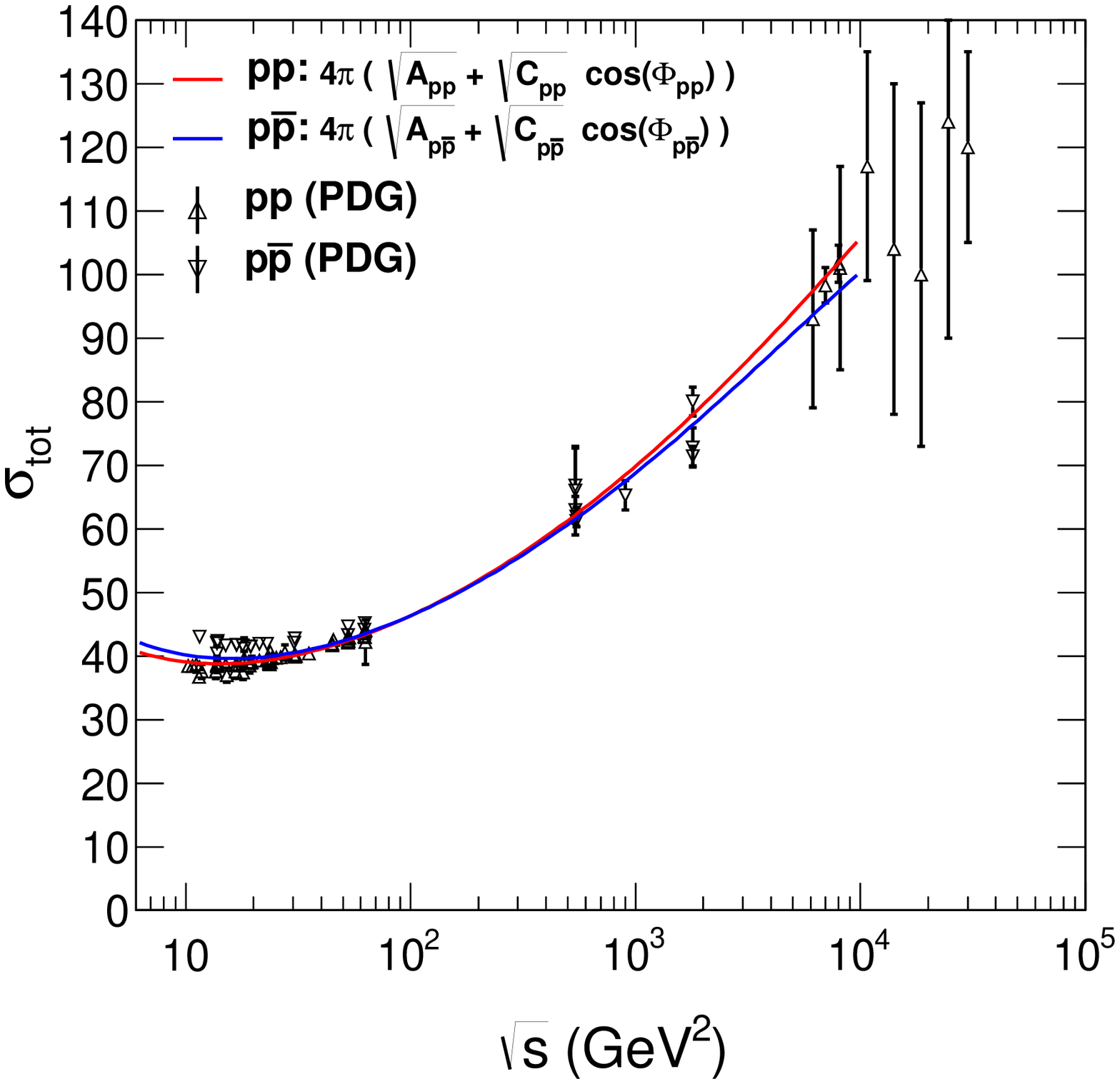}%
\caption{Energy-dependent values of the parameters extracted from a fit to $pp$ and $p \bar p$ data. 
The data on $\sigma_{tot}$ are from the Particle Data Group database~\cite{Beringer:1900zz}.}
\label{fig:pp_params1}%
\end{figure}

The lowest right icon in Fig. \ref{fig:pp_params1} is a ``cross-check", showing the $pp$ and $\bar pp$ total cross sections calculated from Eq. (\ref{norm1}) with the explicit values of the parameters defined by (\ref{AC}), (\ref{BD}) and (\ref{phi}). The $\bar pp$ total cross section turns down at highest energies, deflecting dramatically from that of $pp.$ The reason for that nonphysical effect is the scarcity of $p\bar p$ data, leaving too much freedom in the high-energy extrapolation of the cross section, where one expects asymptotic equality $\sigma_t^{p\bar p}= \sigma_t^{pp}$ for
$s\rightarrow \infty$; see Eq. (47) in Ref. \cite{Wall:1988pa}. This deficiency should, and can, be cured by imposing an additional constraint on the model. This will be done in the next subsection, by fixing (tuning) the parameter $\epsilon_{a_2}$ (the leading powers in $s$ of $\sqrt A$) to be the same in $pp$ and $\bar pp$ scattering.

\clearpage

\subsection{Step 3: Tuning (refitting) the parameters by imposing the asymptotic constraint $\sigma_t^{p \bar p}= \sigma_t^{pp},$
$s\rightarrow \infty$}

\begin{figure}\label{4}%
\includegraphics[width=7cm]{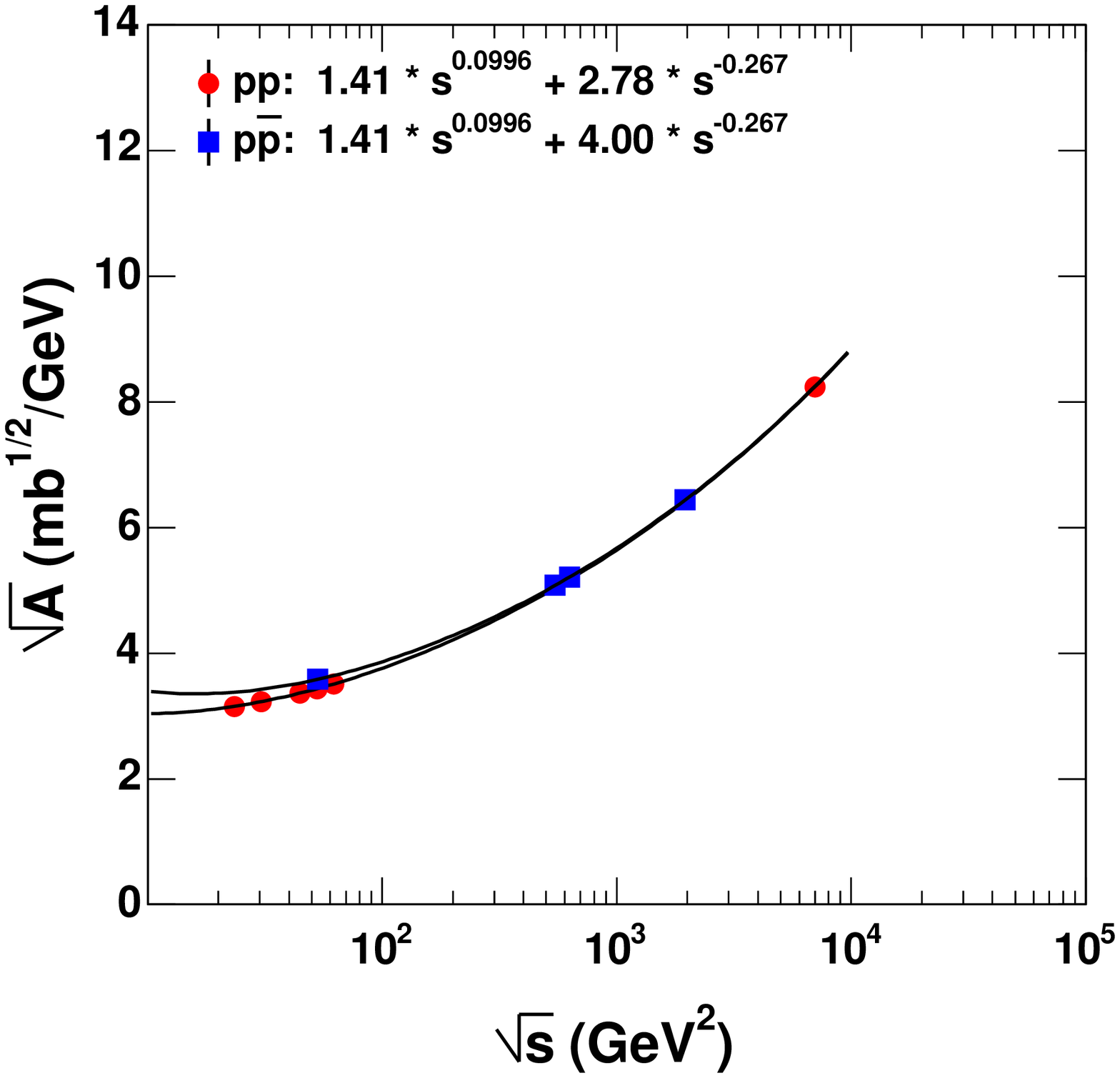}%
\includegraphics[width=7cm]{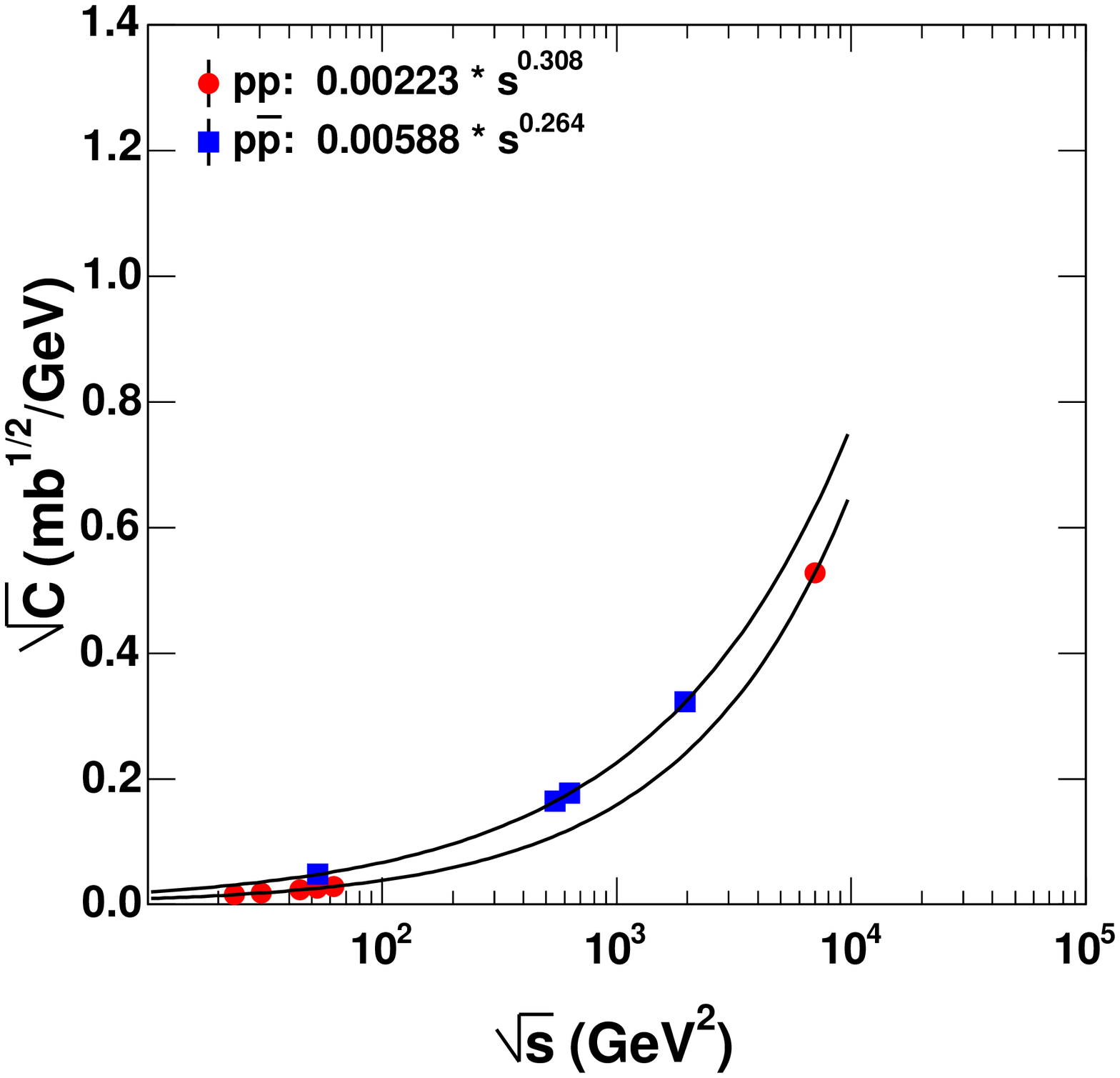}%
\linebreak
\includegraphics[width=7cm]{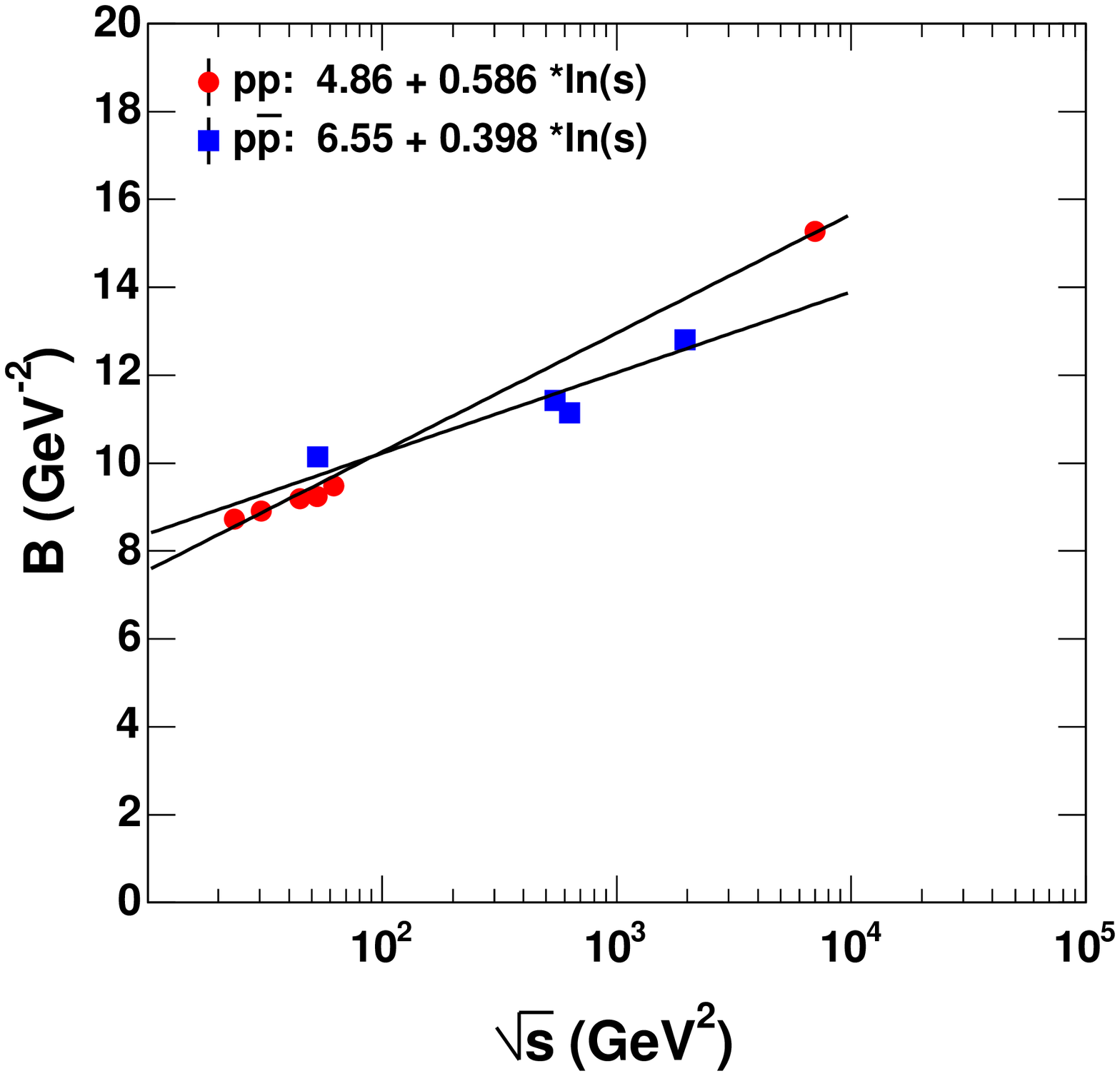}%
\includegraphics[width=7cm]{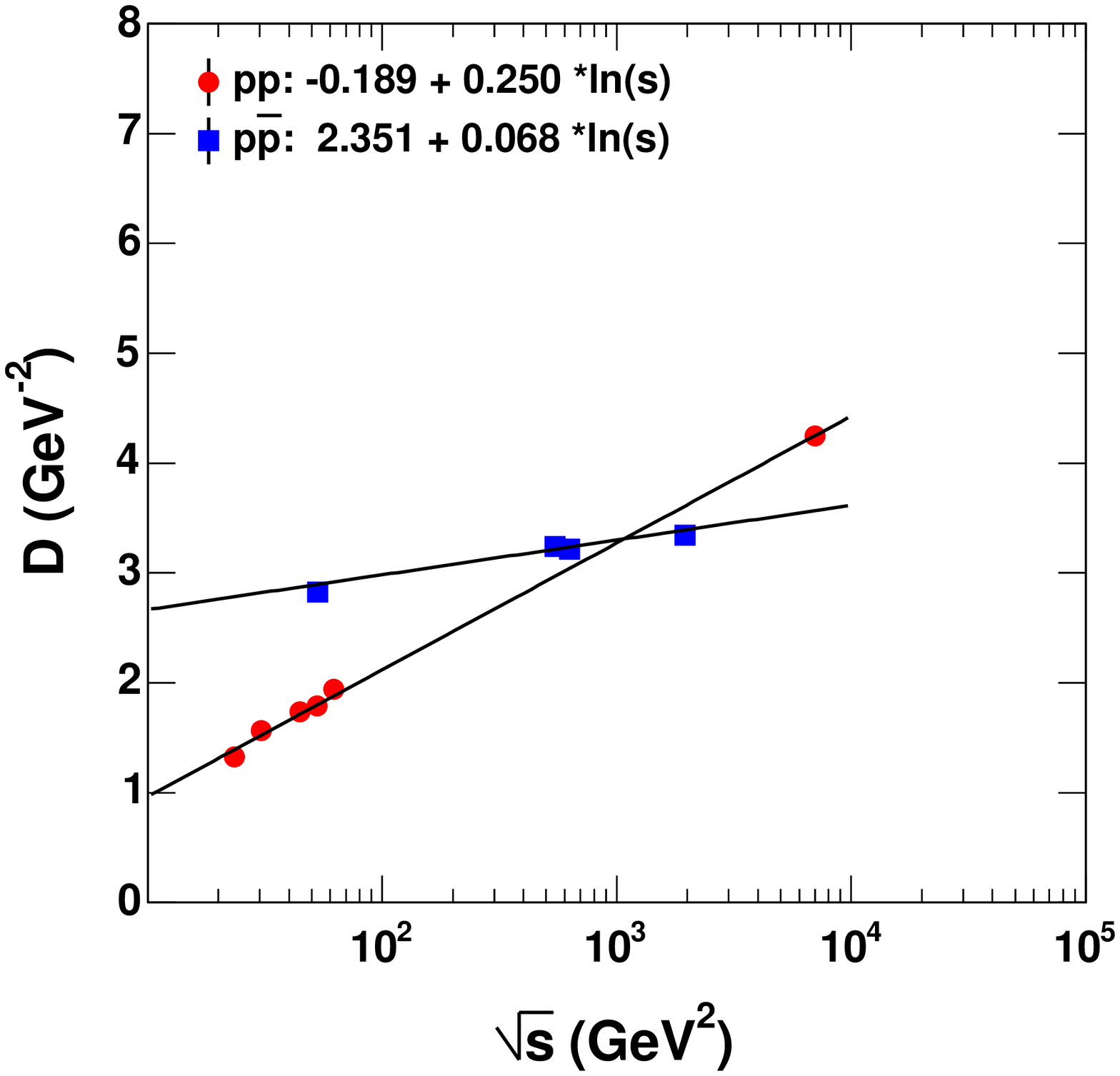}%
\linebreak
\includegraphics[width=7cm]{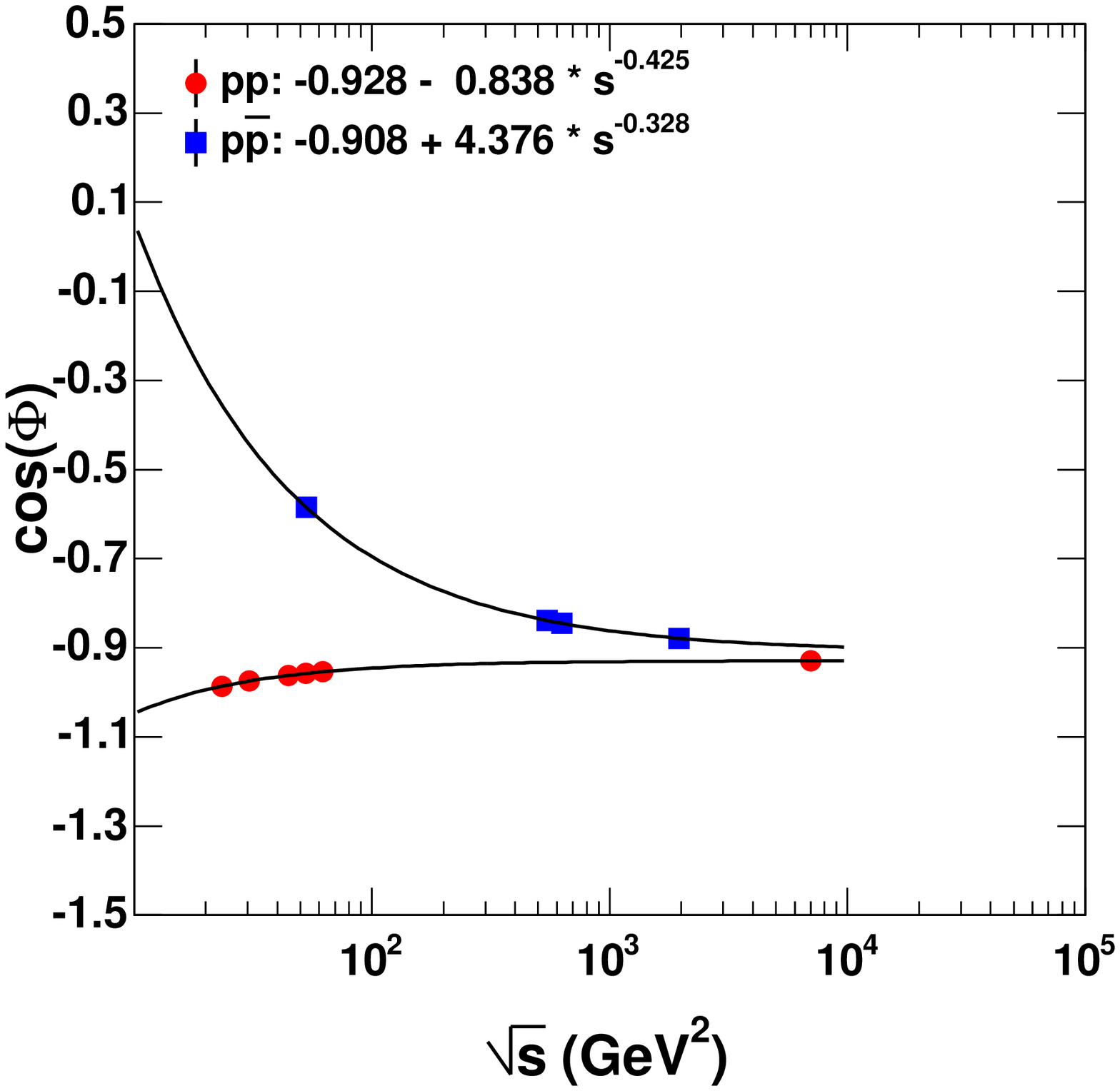}%
\includegraphics[width=7cm]{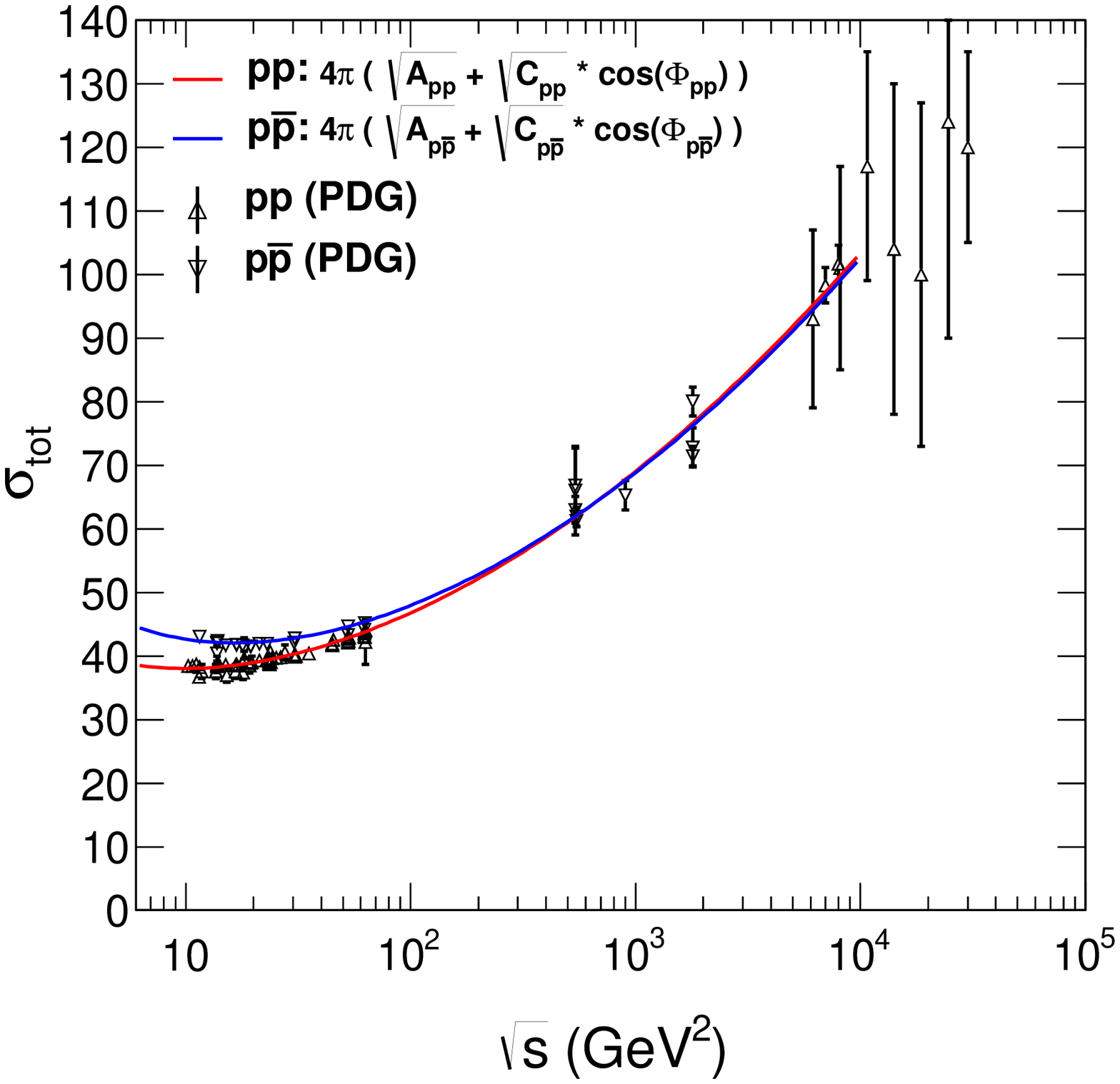}%
\caption{Energy-dependent values of the parameters from a fit to $pp$ and $p\overline{p}$ data, constrained by $\sigma_t^{p \bar p}= \sigma_t^{pp}$ as $\sqrt{s} \rightarrow \infty$. The lowest right icon shows a cross-check for the (asymptotically converting) total cross sections.
}%
\label{fig:pp_params2}%
\end{figure}

The above, unbiased fit does not satisfy automatically the required (see \cite{Wall:1988pa}) 
asymptotic constraint $\sigma_t^{pp}=\sigma_t^{\bar pp}$ since the available freedom (especially 
due to the lack of simultaneous $pp$ and  $p\bar p$ elastic scattering data at $\sqrt{s}=$ 540, 630,
1800 and 7000 GeV) leaves much freedom for the extrapolation to energies beyond the existing accelerators. To remedy this problem, we have tuned the parameters to meet the above constraint in the currently available energy range.  Below are the results of the ``tuned" fit (see Fig. \ref{fig:pp_params2}) satisfying the asymptotic condition 
$\sigma_t^{pp}=\sigma_t^{\bar pp}$ in the $\sqrt{s}\le $ 14 TeV energy range.

The refitted $s$-dependent values of the parameters for $pp$ and $p\overline{p}$ scatterings are

\begin{eqnarray}
\sqrt{A_{pp}(s)}&=& 1.41 s^{0.0966} + 2.78 s^{-0.267}, \\
\sqrt{C_{pp}(s)}&=& 0.00223 s^{0.308},\nonumber\\
B_{pp}(s)&=&4.86 + 0.586 \ln s,\nonumber \\
D_{pp}(s)&=&-0.189 + 0.250 \ln s. \nonumber \\
cos(\phi_{pp}(s))&=& -0.928 - 0.838 s^{-0.425}. \nonumber
\label{eq:PBfunc_pp2}
\end{eqnarray}

\begin{eqnarray}
\sqrt{A_{p\overline{p}}(s)}&=& 1.41 s^{0.0996} + 4.00 s^{-0.267}, \nonumber\\
\sqrt{C_{p\overline{p}}(s)}&=& 0.00588 s^{0.264},\nonumber\\
B_{p\overline{p}}(s)&=&6.55 + 0.398 \ln s,\nonumber \\
D_{p\overline{p}}(s)&=& 2.351 + 0.068 \ln s, \nonumber \\
cos(\phi_{p\overline{p}}(s))&=& -0.908 + 4.376 s^{-0.328}. \nonumber
\label{eq:PBfunc_ppa2}
\end{eqnarray}

\clearpage
\clearpage

\section{The Odderon}\label{Odderon}
The existence of a parametrization for both $pp$ and $\bar pp$ scattering offers the possibility 
to extract the odd-$C$ contribution by using the formula
\begin{equation} \label{odd}
{\cal A}_{pp}^{\bar pp}={\cal A}_{even}\pm {\cal A}_{odd},
\end{equation} 
where ${\cal A}_{even}$ and ${\cal A}_{odd}$ are, respectively, the $C$-even and $C$-odd components of the scattering amplitude; see Table 1 in Ref. \cite{Jenkovszky:2011hu}.

\begin{figure}\label{5}%
\includegraphics[width=7cm]{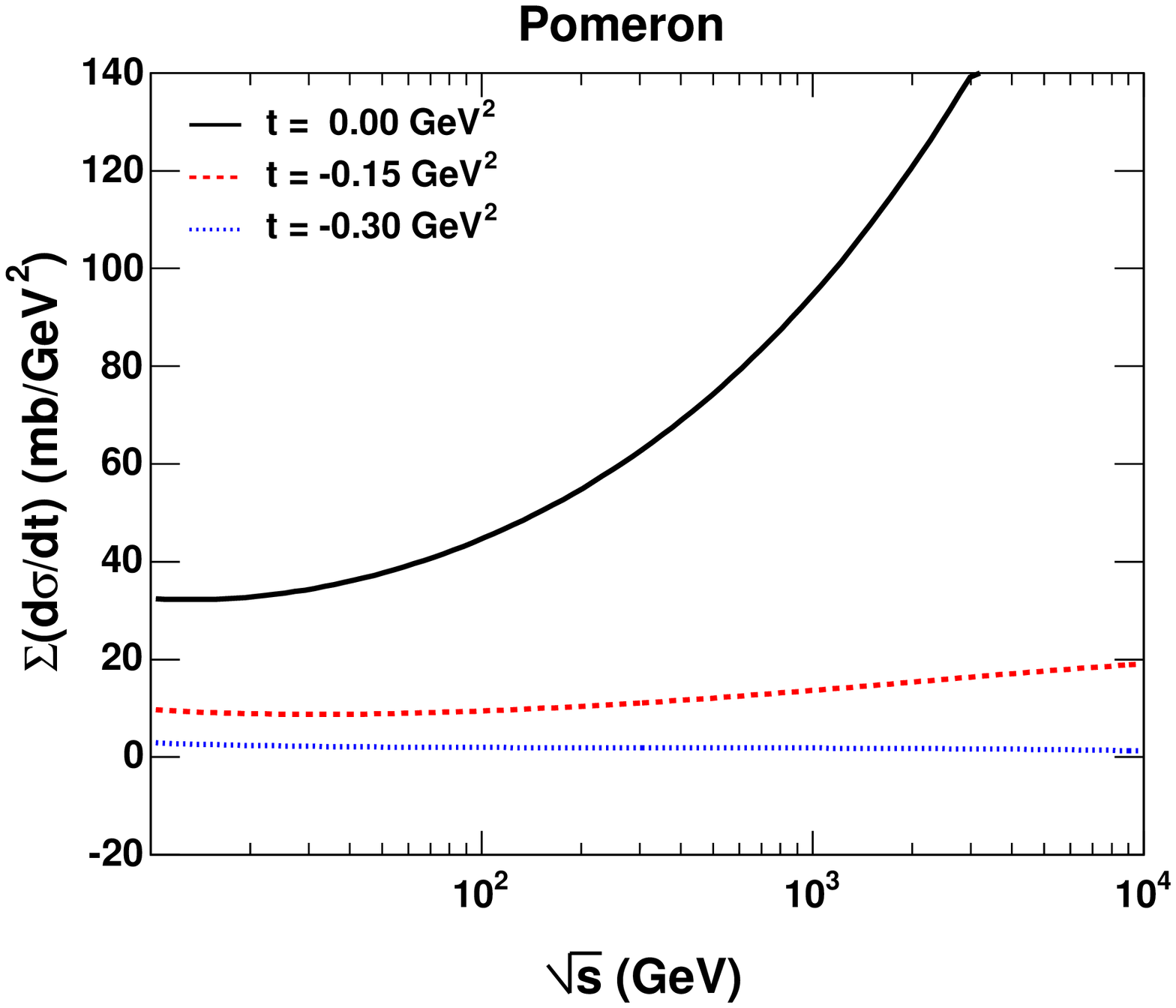}%
\includegraphics[width=7cm]{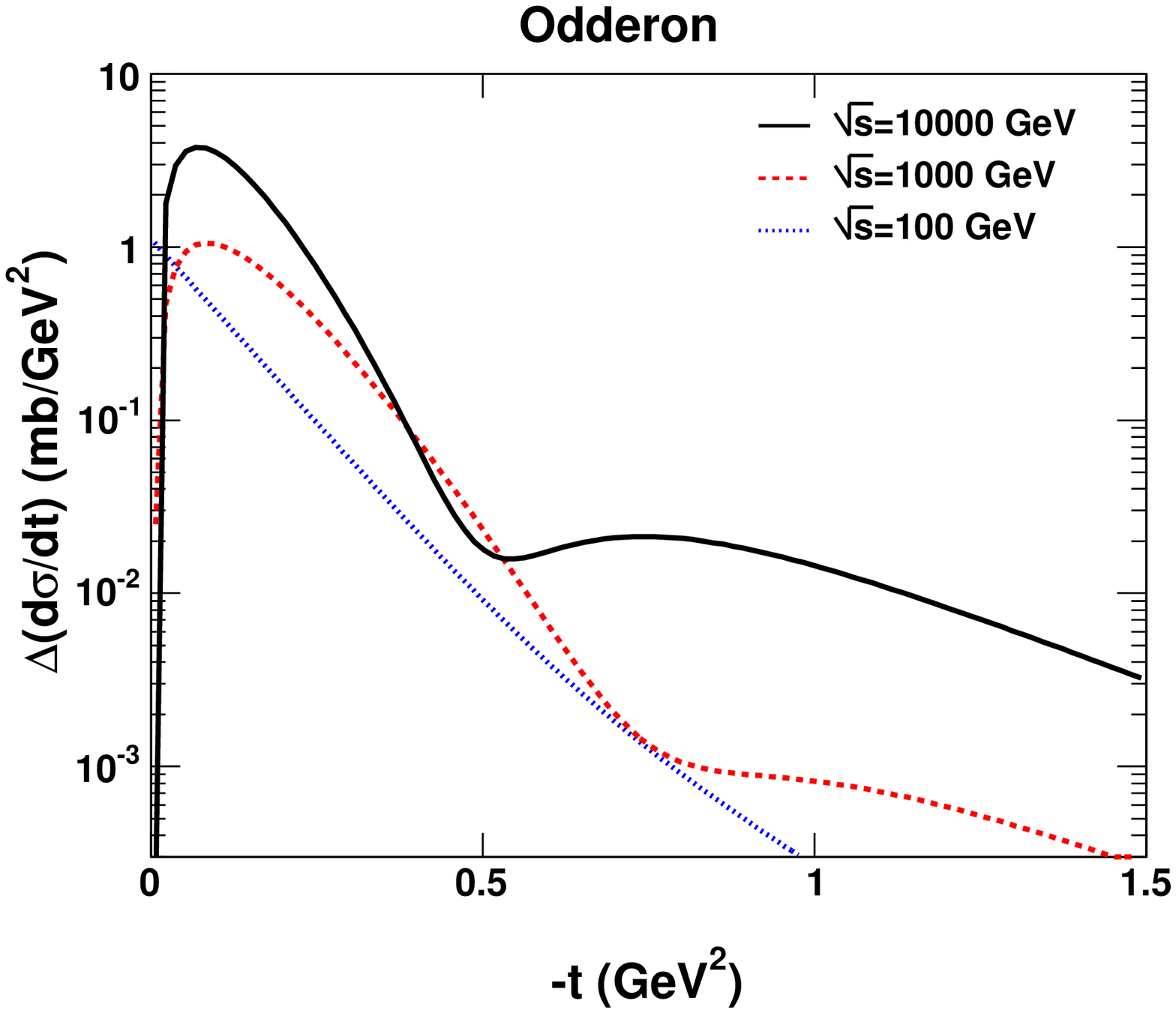}%
\linebreak
\includegraphics[width=7cm]{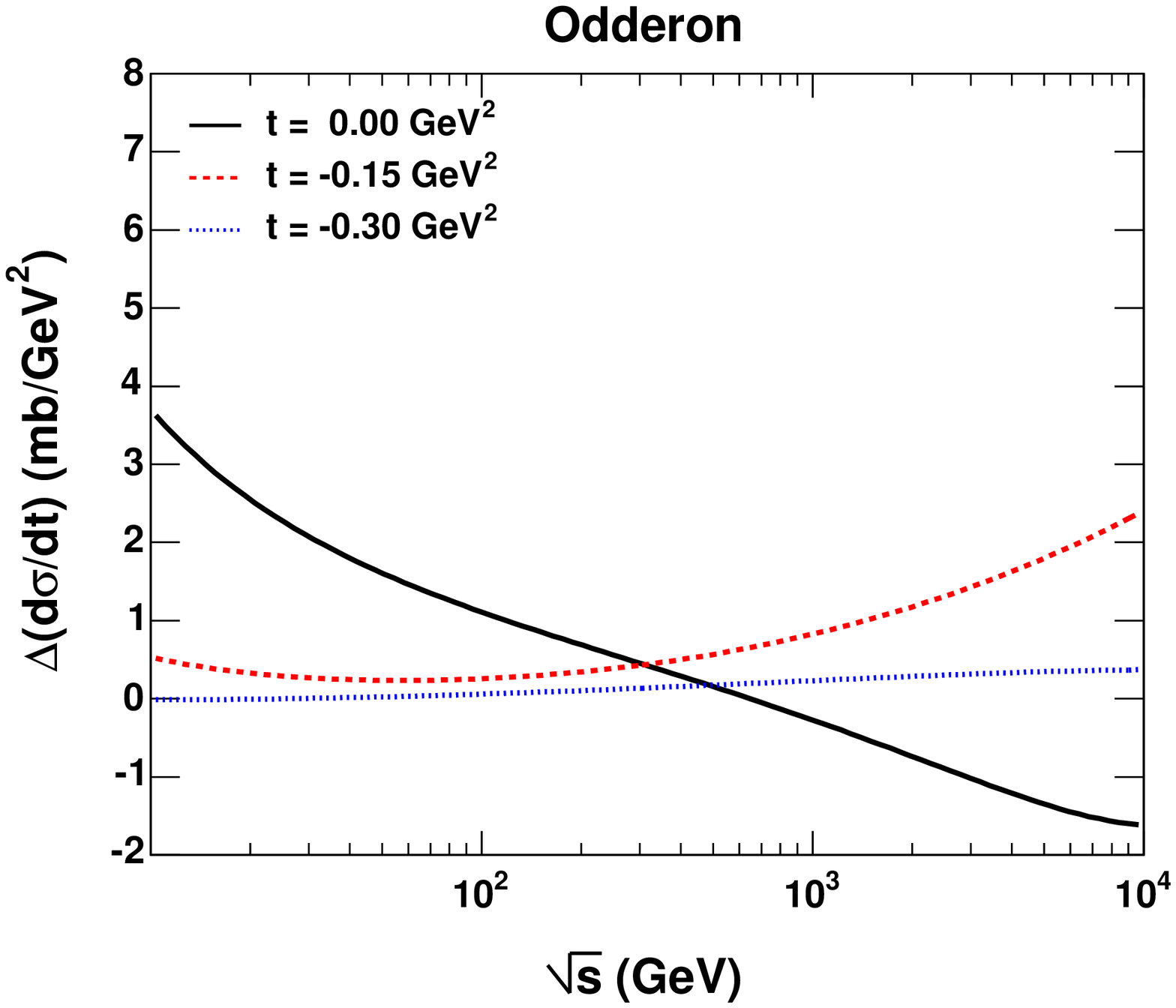}%
\includegraphics[width=7cm]{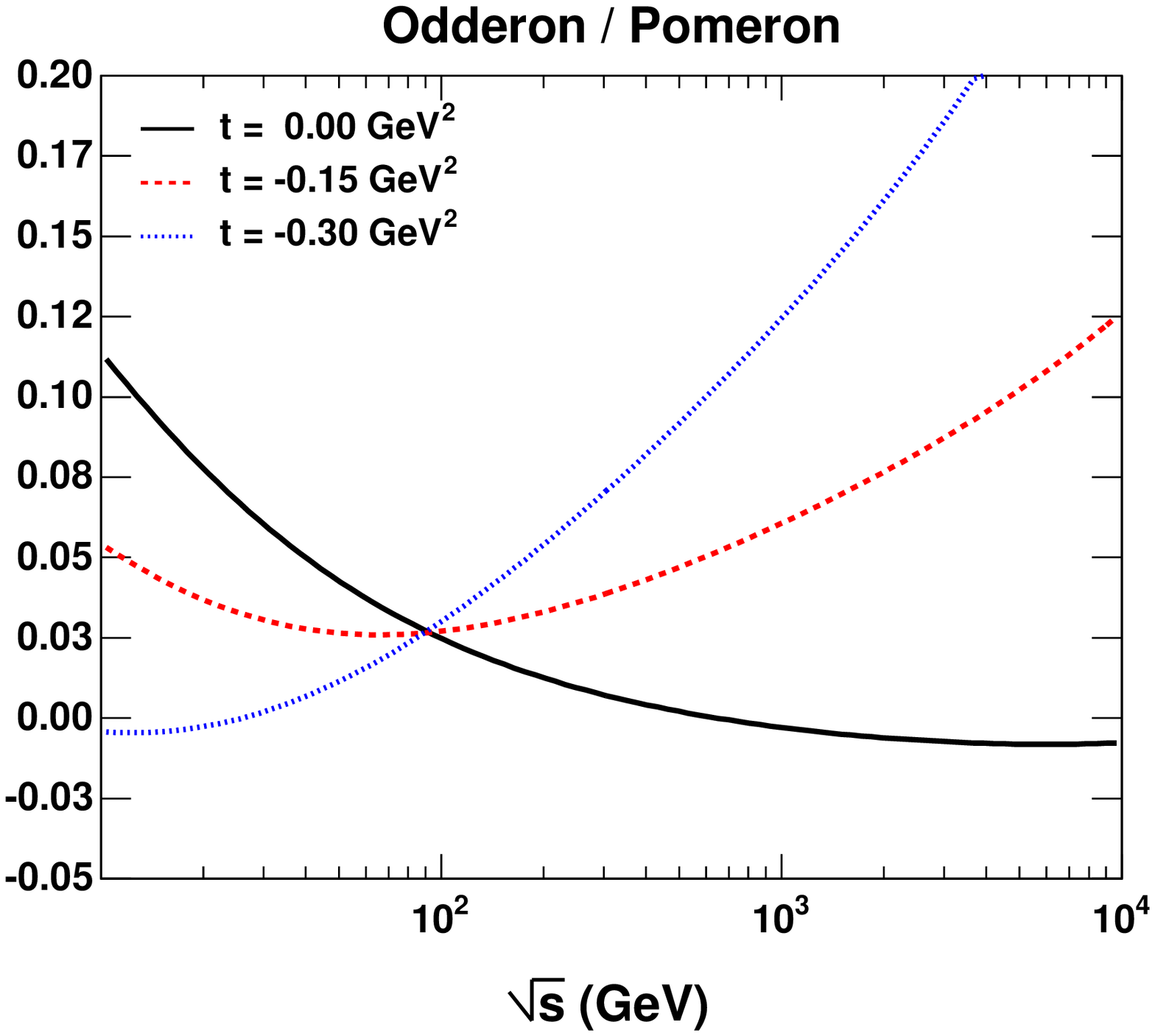}%
\caption{Odd (Odderon) and even (Pomeron) parts of the $d\sigma/dt$ cross sections calculated
from the difference $|{\cal A}|^2_{\bar pp}- |{\cal A}|^2_{pp}=\Delta_{Odd}$ and 
from the sum       $|{\cal A}|^2_{\bar pp}+ |{\cal A}|^2_{pp}=\Sigma_{Pom}$
fitted to the data.}%
\label{fig:odderon1}%
\end{figure}

\begin{figure}\label{Odd}%
\includegraphics[width=10cm]{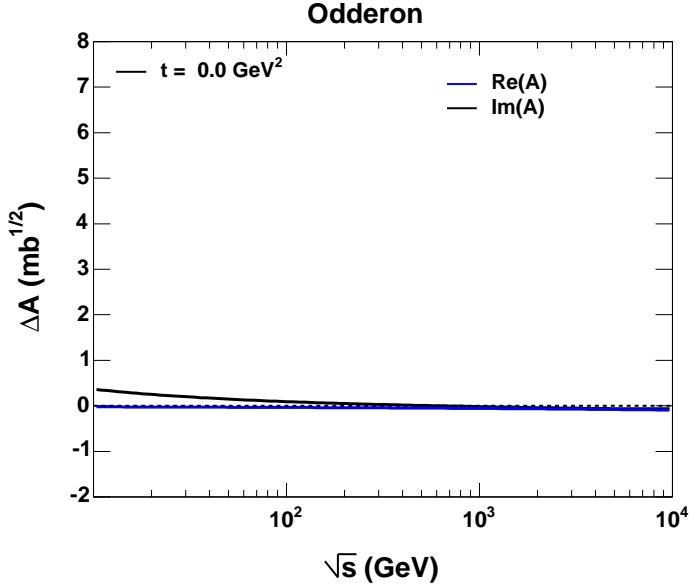}%
\caption{Real and imaginary parts of the difference ${\cal A}^{\bar pp}-
{\cal A}^{pp}=\Delta{\cal A}$ calculated from the present model fitted to the data. As seen from the figure, this difference tends to a (small) constant, corresponding to a unit intercept Odderon \cite{BLV}.}  
\label{fig:odderon2}%
\end{figure}

While 
the $C$-even component contains the Pomeron and the $f$ trajectory (both known), the $C$-odd part is made of the poorly known Odderon and the familiar $\omega$ trajectory. 
At the  LHC energies, the contribution from secondary trajectories, e.g., $f$ and $\omega$, is negligible (see Ref. \cite{Jenkovszky:2011hu}); therefore, by taking the difference between the  known (fitted) $p\bar p$ and 
$pp$ amplitudes, one gets a pure odd-$C$ contribution that, in the LHC energy range, is the Odderon. 
From the explicit expressions for $pp$ and $\bar pp$ amplitudes (cross sections), we calculate the Odderon amplitude (or its contribution to the cross section) by taking the difference ${\cal A}^{\bar pp}-
{\cal A}^{pp}={\cal A}_{Odd}.$
The result (the energy dependence for several fixed values of 
$t$ and $t$ dependence for several fixed values of $s$) is shown in Fig. 4.

The extracted model parameters are used then to evaluate the even and odd contributions to the
forward scattering amplitude. In Fig. \ref{fig:odderon1} we show the Pomeron and the Odderon contributions as the sum or the difference of the differential cross section of $p\bar p$ and $pp$ elastic scattering. We see that, as expected, the Pomeron dominates at large colliding energies, while the Odderon contribution is small and at $t=0$ even changes sign. A particularly interesting feature is shown on the lower right panel of Fig. \ref{fig:odderon1}, where the Odderon/Pomeron ratio is shown at different values of $t$ at various $\sqrt{s}$. Apparently, at $\sqrt{s}\approx 100$ GeV,
the Odderon/Pomeron ratio becomes $t$ independent and the t-dependent curves pass through the same point of
about 0.03.

An important finding of our paper is the near energy independence of the odd-$C$ contribution to the scattering amplitude, which, at high (e.g., those of the LHC) energies, is dominated by the Odderon exchange. In Fig. \ref{fig:odderon2} we plot the odd-$C$ contribution calculated from the difference (\ref{odd}). Both the real and imaginary parts of this difference at $t=0$ tend to a (small) constant value, which correspond to a unit-intercept Odderon as predicted  \cite{BLV} from quantum chromodynamics.    

\section{Conclusions}
The present work is a semiquantitative estimate of the possible odd-$C$ contribution to the scattering amplitude (cross sections) at high energies.
``Semiquantitative" implies limitations due to the following:
 
\begin{itemize}

\item For the sake of simplicity, we ignore the low-$t$ nonexponential behavior (sharpening) of the differential cross section. This simplification has dramatic impact on the 
low-$t$ behavior of the Odderon contribution because of the large errors due to the cancellation of the Pomeron contribution. The importance of the low-$t$ effects was emphasized, e.g., in Refs. \cite{Grau:2012wy, Fagundes:2013aja}. 
 
\item The $s$- (rather than $t$-) dependent signature factor (phase) is in agreement with the data, but it is in contrast to expectations based on Regge phenomenology.

\item The odd-$C$ contribution to the amplitude, equal to the difference (\ref{odd}), shown in Fig. \ref{fig:odderon2}, tends to a constant limit determined by a unit-intercept Odderon, predicted by Bartels, Lipatov, and Vacca \cite{BLV}.

\item It is an oversimplified treatment of the low-energy (``secondary Reggeons") contributions (Odderon and Pomeron here implies, generally, odd and even exchanges). 

\item There is the absence, for the moment, of any physical interpretation in terms of Reggeon exchanges of the components in the PB ansatz.  
\end{itemize}   

Given these limitations and simplifications, our approach can be considered as semiquantitative, showing however some new aspects of the enigmatic Odderon.  

We strongly recommend to run the LHC accelerator at the injection energy $\sqrt{s} = 900$ GeV and at the
Tevatron energy of $1.8$-$1.96$ TeV, so that the missing energy range of $pp$ elastic scattering will be covered and elastic
$pp$ scattering data will be measured in the region where elastic $p\bar p$ scattering is already measured.
It would also be desireable to measure elastic $pp$ scattering at the 
$\sqrt{s} = 500$ GeV region --- the upper energy range of the RHIC accelerator.

Using the presently available data, the indefinite rise of the $C(s)$,
multiplied by a negative ``signature factor" $\cos\phi,$ prevents the use of the generalized PB model 
beyond the LHC energy region.

In the future we intend to
\begin{itemize}

\item rewrite the PB ansatz with correct, $t$-dependent Regge signature factors \cite{Jenkovszky:2014bwa}, remembering that in the present study the even and odd parts of the amplitude differ only by the values of the fitted parameters, without the identification of particular Regge exchanges (trajectories);   

\item fit to the data to provide hints concerning the physical meaning of the components (two or more);
for example, our fits (and those of Refs. \cite{Phillips:1974vt, Grau:2012wy, Fagundes:2013aja, Fagundes:2013cja}) indicate that $A>>C,$ but $C$ rises with energy faster 
than $A$. Similar considerations may help in relating the second term with the
Odderon or a ``hard" Pomeron, for which the slopes ($B$ and $D$) may be indicative. 

\item add lower-lying contributions (subleading Reggeons) which are inevitable at lower energies, at the prise of giving up of the attractive simplicity of the model;

\item use other asymptotic Pomeranchuk-like constraints in determining the parameters (see
the work in Ref. ~\cite{Pancheri:2014roa} that points to this interesting direction); 

\item calculate and fit the model to other observables, e.g., the ratio of the real to imaginary parts of the amplitude $\rho(s,t)$, of the slope $B(s,t)$, etc. Its relation to inelastic processes (see \cite{Goulianos:1982vk}) may offer additional information.   

\end{itemize}

\begin{acknowledgments}
L.J. acknowledges cordial hospitality by the Wigner RCP, Budapest, Hungary, where this work was completed. He is also grateful to D.A. Fagundes and G. Pancheri for useful discussions and correspondence.   
This work was partially supported by the Hungarian OTKA Grant No. NK101438.
\end{acknowledgments}

\end{document}